\documentclass[aps,prc,showkeys,nofootinbib]{revtex4}

\usepackage{color} 

\usepackage{graphicx}
\usepackage{dcolumn}
\usepackage{amsthm}

\usepackage{isotope}
\usepackage{braket}
\usepackage{physics}

\begin{document}

\title{Synthesis of elements in compact stars in pycnonuclear reactions with Carbon isotopes:
Quasibound states versus states of zero-points vibrations}

\author{Sergei~P.~Maydanyuk$^{(1,2)}$}\email{sergei.maydanyuk@wigner.hu}%
\author{Gyorgy~Wolf$^{(1)}$}\email{wolf.gyorgy@wigner.hu}%
\author{Kostiantyn~A.~Shaulskyi$^{(2)}$}\email{shaulskyi@kinr.kiev.ua}%

\affiliation{$^{(1)}$Wigner Research Center for Physics, Budapest, 1121, Hungary}
\affiliation{$^{(2)}$Institute for Nuclear Research, National Academy of Sciences of Ukraine, Kyiv, 03680, Ukraine}

\date{\small\today}



\begin{abstract}
\textbf{(1) Purpose:}
Conditions of formation of compound nuclear system needed for synthesis of heavy nuclei
in pycnonuclear reactions in compact stars
are studied
on a quantum mechanical basis.
%
\textbf{(2) Methods:}
Method of multiple internal reflections is generalized for pycnoreactions in compact stars with
new calculations of quasibound spectra and spectra of zero-point vibrations.
%
\textbf{(3) Results:}
Peculiarities of the method are analyzed for reaction with isotopes of Carbon. 
The developed method takes into account continuity and conservation of quantum flux (describing pycnonuclear reaction)
inside the full spacial region of reaction including nuclear region.
This gives appearance of new states (called as quasibound states), in which compound nuclear systems of Magnesium are formed with the largest probability.
These states have not been studied yet 
in synthesis of elements in stars.
Energy spectra of zero-point vibrations and spectra of quasibound states are estimated with high precision for reactions with isotopes of Carbon.
%
%
At the first time influence of plasma screening on quasibound states and states of zero-point vibrations in pycnonuclear reactions has been studied.
\textbf{(4) Conclusion:}
The probability of formation of compound nuclear system in quasibound states in pycnonuclear reaction is essentially larger than
the probability of formation of this system in states of zero-point vibrations
studied by Zel'dovich and followers.
So, synthesis of Magnesium from isotopes of Carbon is more probable through the quasibound states
than through the states of zero-point vibrations in compact stars.
Energy spectra of zero-point vibrations are changed essentially after taking plasma screening into account.
Analysis shows that from all studied isotopes of Magnesium only \isotope[24]{Mg} is stable after synthesis
at energy of relative motion of 4.881~MeV of incident nuclei \isotope[12]{C}.
\end{abstract}

\keywords{
pycnonuclear reaction,
compact star,
neutron star,
multiple internal refleclections,
coefficients of penetrability and reflection,
fusion,
quasibound state,
energy of zero-point vibrations,
compound nucleus,
dense nuclear matter,
tunneling
}

\maketitle


\section{Introduction
\label{sec.introduction}}


Pycnonuclear burning occurs in dense and cold cores of white dwarfs~\cite{Salpeter_VanHorn.1969.AstrJ} and
crusts of accreting neutron stars~\cite{Schramm.1990.AstrJ,Haensel.1990.AstronAstrophys}.
Such a phenomenon known as pycnonuclear reaction~\cite{Cameron.1959b.AstrJ} is reaction at sufficiently high densities in stars
where zero-point vibrations of nuclei in the lattice sites leads to an essential increasing rate of formation of more heavy nuclei.
Insight to this phenomenon was given by Zel'dovich
who estimated zero-point energy as in the ground state of discrete energy spectrum of the harmonic oscillator potential which is formed near middle point between two nuclei in lattice sites~\cite{Zeldovich.1965.AstrJ}.
Rates of reactions derived at such zero-point energies are estimated for some atomic nuclei in lattices in compact stars~\cite{ShapiroTeukolsky.2004.book}.

Key process in pycnonuclear reactions is fusion of nuclei producing new nucleus with larger mass.
In Ref.~\cite{Yakovlev.2006.PRC} this question was analyzed for reactions with nuclei of different charges and masses,
where authors derived the astrophysical $S$-factors for carbon-oxygen and oxygen-oxygen fusion reactions on the microscopic basis.
Phenomenological expressions for reaction rates containing several fit parameters were found. 
In Ref.~\cite{Afanasjev.2010.ADNDT} the astrophysical $S$-factors were derived for
946 fusion reactions involving stable and neutron-rich isotopes of C, O, Ne, and Mg for center-of-mass energies from 2 to $\approx 18$--30~MeV.
Results in that paper can be converted to thermonuclear or pycnonuclear reaction rates to simulate
stellar burning at high temperatures and nucleosynthesis in high density environments.
Large collection of astrophysical $S$ factors and their compact representation for
isotopes of Be, B, C, N, O, F, Ne, Na, Mg, and Si
were presented in Ref.~\cite{Afanasjev.2012.PRC}.
Database of $S$ factors was created for about 5000 nonresonant fusion reactions in those researches.
Structure of the multi-component matter (a regular lattice, a uniform mix, etc.) in these reactions,
plasma screening \cite{Kravchuk.2014.PRC},
rates of reactions in a wide range of temperatures and stellar densities~\cite{Gasques.2005.PRC,Yakovlev.2006.PRC}
have been studied by many researchers.

It has been known that cross sections of reactions are changed essentially after taking into account conservation of quantum fluxes
in the internal region of nuclear system~\cite{Maydanyuk.2011.JPS,Maydanyuk.2015.NPA,Maydanyuk_Zhang_Zou.2017.PRC}.
This question has been studied in details for $\alpha$ decays of nuclei and captures of $\alpha$-particles by nuclei.
For example, the study of captures of $\alpha$ particles by nuclei~\cite{Maydanyuk.2015.NPA,Maydanyuk_Zhang_Zou.2017.PRC}
shows the importance to take into account internal nuclear processes during capture 
that depends on the shape of the nuclear potential.
After such an inclusion, different scenarios of capture (before fusion) appear,
those processes give resulting cross sections with difference up to 4 times at the same beam energies of $\alpha$ particles used in experiments.
Such changes are controlled by additional independent parameters appeared from fully quantum study.
At the same time, such processes are completely missed in the WKB-approximation where such quantum parameters do not exist.
It is important to note that this dependence of cross sections in fully quantum study is not small.
For example, it can be larger essentially than inclusion of nuclear deformations to the calculation of cross sections without such quantum parameters.
Up to now, method in Ref.~\cite{Maydanyuk.2015.NPA} is the most accurate approach in description of experimental data for the $\alpha$-capture
(this calculation is 
in Fig.~3~(b) in Ref.~\cite{Maydanyuk_Zhang_Zou.2017.PRC}
for $\alpha + \isotope[44]{Ca}$ in comparison with experimental data~\cite{Eberhard.1979.PRL}).

In the fully quantum study the accuracy in determination of penetrability of barrier and cross section is about of $10^{-14}$,
while such an accuracy in the WKB-approximation is about $10^{-1}$--$10^{-3}$~\cite{Maydanyuk.2015.NPA,Maydanyuk_Zhang_Zou.2017.PRC}.
Energies for the pycnonuclear reactions are low and there are only processes of deep tunneling under the barrier where the semiclassical approximation is not applicable.
This reinforces our interest to this problem.
These quantum effects have not been studied yet by other researchers for pycnonuclear reactions in stars. 
In Ref.~\cite{Maydanyuk_Shaulskyi.2022.EPJA} investigation of these questions on the fully quantum basis was initiated on example of reaction of $\isotope[12]{C} + \isotope[12]{C}$.
Interest to that reaction is explained by its
impact on nucleosynthesis, energy production and other questions in stellar evolution~\cite{Gasques.2005.PRC,Chien.2018.PRC}.
Also this reaction has a significant impact on the evolution and structure of massive stars with $M \ge M_{\odot}$ ($M_{\odot}$ is the Solar mass).
$\isotope[12]{C} + \isotope[12]{C}$ fusion is known as pycnonulear reaction that reignites carbon-oxygen white dwarf into type Ia supernova explosion.
However, it could be useful to obtain more complete picture for the systematic analysis for nuclear processes and fusion for reactions with isotopes of Carbon.
So, in this paper we perform such an investigation for pycnonuclear reactions with Carbon.

The paper is organized in the following way.
In Sec.~\ref{sec.3.2} a new generalized formalism of the multiple internal reflections is reviewed with focus on new elements for fusion and quasibound states in pycnonuclear reactions.
In Sec.~\ref{sec.analysis} reactions with isotopes of Carbon on the basis of the method are studied,
with calculations of penetrabilities of the potential barriers, probabilities of formation of the compound nucleus,
estimation of energies for zero-point vibrations and quasibound states, etc..
In Sec.~\ref{sec.screening} influence of plasma screening on properties of the pycnonuclear reaction is studied on example
of $\isotope[12]{C} + \isotope[12]{C}$.
Conclusions are summarized in Sec.~\ref{sec.conclusion}.

\section{Method of multiple internal reflections in description of nucleus-nucleus scattering with possibility of fusion
\label{sec.3.2}}

%
Let us consider a process of capture of the particle by the nucleus with the radial barrier of arbitrary shape,
which has successfully been approximated by a sufficiently large number $N$ of rectangular steps:
\begin{equation}
  V(r) = \left\{
  \begin{array}{cll}
    V_{1}   & \mbox{at } r_{\rm min} < r \leq r_{1}      & \mbox{(region 1)}, \\
    \ldots   & \ldots & \ldots \\
    V_{N_{\rm cap}}   & \mbox{at } r_{N_{\rm cap} - 1} \leq r \leq r_{\rm cap}         & \mbox{(region $N_{\rm cap}$)}, \\
    \ldots   & \ldots & \ldots \\
    V_{N}   & \mbox{at } r_{N-1} \leq r \leq r_{\rm max} & \mbox{(region $N$)},
  \end{array} \right.
\label{eq.3.2.1.1}
\end{equation}
%
%
%
%
where $V_{j}$ are constants ($j = 1 \ldots N$). %
We denote the first region with a left boundary at point $r_{\rm min}$.
In addition to study in Ref.~\cite{Maydanyuk.2015.NPA},
in this paper we shall assume that the capture of the particle by the nucleus takes place most probably in region with number $N_{\rm capture}$
after its tunneling through the barrier,
and further propagation of waves inside the internal region exists and should be studied.
This logic follows from the condition of continuity of fluxes in quantum mechanics, that is strict condition, and it should be included to formalism.
Solution of the radial wave function (up to its normalization) for the above barrier energies has the following form:
\begin{equation}
\chi(r) = \left\{
\begin{array}{lll}
   \alpha_{1}\, e^{ik_{1}r} + \beta_{1}\, e^{-ik_{1}r},
     & \mbox{at } r_{\rm min} < r \leq r_{1} & \mbox{(region 1)}, \\
   \alpha_{2}\, e^{ik_{2}r} + \beta_{2}\, e^{-ik_{2}r},
     & \mbox{at } r_{1} \leq r \leq r_{2} & \mbox{(region 2)}, \\
   \ldots & \ldots & \ldots \\
   \alpha_{N-1} e^{ik_{N-1}r} + \beta_{N-1} e^{-ik_{N-1}r}, &
     \mbox{at } r_{N-2} \leq r \leq r_{N-1} & \mbox{(region $N-1$)}, \\
   e^{-ik_{N}r} + A_{R}\,e^{ik_{N}r}, & \mbox{at } r_{N-1} \leq r \leq r_{\rm max} & \mbox{(region $N$)},
\end{array} \right.
\label{eq.3.2.1.2}
\end{equation}
%
%
where $\alpha_{j}$ and $\beta_{j}$ are unknown amplitudes, $A_{R}$ is unknown amplitude of the full reflection,
and $k_{j} = \frac{1}{\hbar}\sqrt{2m(\tilde{E}-V_{j})}$ are complex wave numbers.
We fix the normalization so that the modulus of amplitude of the starting wave $e^{-ik_{N}r}$ equals to unity.
We shall find a solution of this problem by the multiple internal reflections approach.
This method was described in details in Ref.~\cite{Maydanyuk.2015.NPA}
(see also Refs.~\cite{Maydanyuk.2002.JPS,Maydanyuk.2011.JPS,Maydanyuk_Zhang_Zou.2017.PRC,Maydanyuk.2000.UPJ}).
In this paper we will indicate only the main aspects of this approach in order to explain logic of obtaining solution.

Each wave in an arbitrary step can be represented as contribution of the exponential factor $e^{\pm i\,k_{j}\,r}$ with unknown amplitude.
The amplitude of the wave transmitted through the boundary with number $j$ is written down as the product of the amplitude of the corresponding wave incident on this boundary and the new factor $T_{j}^{\pm}$
(i.~e., the amplitude of transition through the boundary with number $j$).
The bottom index indicates the number of the boundary, upper sign ``$+$'' or ``$-$'' is the direction of the incident wave to the right or left, respectively.
The amplitude of the reflected wave from the boundary with number $j$ is associated with the amplitude of the wave incident on this boundary via new factors $R_{j}^{\pm}$.
The coefficients
$T_{1}^{\pm}$ \ldots $T_{N-1}^{\pm}$ and $R_{1}^{\pm}$ \ldots $R_{N-1}^{\pm}$
are found from the recurrence relations as
(see Ref.~\cite{Maydanyuk.2011.JMP})
\begin{equation}
\begin{array}{ll}
\vspace{2mm}
   T_{j}^{+} = \displaystyle\frac{2k_{j}}{k_{j}+k_{j+1}} \,e^{i(k_{j}-k_{j+1}) r_{j}}, &
   T_{j}^{-} = \displaystyle\frac{2k_{j+1}}{k_{j}+k_{j+1}} \,e^{i(k_{j}-k_{j+1}) r_{j}}, \\
   R_{j}^{+} = \displaystyle\frac{k_{j}-k_{j+1}}{k_{j}+k_{j+1}} \,e^{2ik_{j}r_{j}}, &
   R_{j}^{-} = \displaystyle\frac{k_{j+1}-k_{j}}{k_{j}+k_{j+1}} \,e^{-2ik_{j+1}r_{j}}.
\end{array}
\label{eq.3.2.1.3}
\end{equation}

Crucial point in the formalism of multiple internal reflections for a barrier with really large number of steps is determination of summed waves for an arbitrary region.
Wave propagating to the left in the region with number $j-1$ is formed after transmission through the boundary at point $r_{j-1}$ of all possible incident waves,
produced after all reflections and transmissions of waves in the right part of potential from this boundary.
Amplitude of this wave is determined as the sum of the amplitudes of all waves incident on the boundary at point $r_{j-1}$ multiplied on factor $T_{j-1}^{-}$ as
%
\begin{equation}
\begin{array}{llll}
  \tilde{T}_{j-1}^{-} & = &
    \tilde{T}_{j}^{-} T_{j-1}^{-}
    \Bigl(1 + \sum\limits_{m=1}^{+\infty} (R_{j-1}^{-} \tilde{R}_{j}^{+})^{m} \Bigr) =
    \displaystyle\frac{\tilde{T}_{j}^{-} T_{j-1}^{-}} {1 - R_{j-1}^{-} \tilde{R}_{j}^{+}}.
\end{array}
\label{eq.3.2.1.4}
\end{equation}
%
Summed reflection amplitude $\tilde{R}_{j}^{+}$ includes
waves transferred through boundary at $r_{j}$, propagated to the right,
then after reflections and transmissions returned back to region with number $j$:
\begin{equation}
\begin{array}{lllll}
  \tilde{R}_{j-1}^{+} & = &
     R_{j-1}^{+} + \displaystyle\frac{T_{j-1}^{+} \tilde{R}_{j}^{+} T_{j-1}^{-}} {1 - \tilde{R}_{j}^{+} R_{j-1}^{-}}.
\end{array}
\label{eq.3.2.1.5}
\end{equation}
Recurrent relations~(\ref{eq.3.2.1.4}) and (\ref{eq.3.2.1.5}) calculate consequently
all amplitudes
$\tilde{R}_{N-2}^{+}$ \ldots $\tilde{R}_{N_{\rm cap}}^{+}$,
and $\tilde{T}_{N-2}^{-}$ \ldots $\tilde{T}_{N_{\rm cap}}^{-}$
if at start we use
\begin{equation}
\begin{array}{cccc}
  \tilde{R}_{N-1}^{+} = R_{N-1}^{+}, & \quad
  \tilde{T}_{N-1}^{-} = T_{N-1}^{-}.
\end{array}
\label{eq.3.2.1.6}
\end{equation}
%
Amplitude of summed wave of reflection combining all waves reflected from the boundary at point $r_{j+1}$ and propagating to the left,
is
\begin{equation}
\begin{array}{lcl}
  \tilde{R}_{j+1}^{-} & = &
    R_{j+1}^{-} +
    \displaystyle\frac{T_{j+1}^{-} \tilde{R}_{j}^{-} T_{j+1}^{+}} {1 - R_{j+1}^{+} \tilde{R}_{j}^{-}}.
\end{array}
\label{eq.3.2.1.8}
\end{equation}
%
Summed amplitude $A_{T, {\rm bar}}$ of transition through the barrier
(summed amplitude $A_{R, {\rm bar}}$ of reflection from the barrier)
is determined via all waves transmitted through (reflected from) the potential region with the barrier from $r_{\rm cap}$ to $r_{N-1}$ as
\begin{equation}
\begin{array}{llllll}
  A_{T, {\rm bar}} = \tilde{T}_{N_{\rm cap}}^{-}, &
  A_{R, {\rm bar}} = \tilde{R}_{N-1}^{-}, &
  {\rm where}\;
  \tilde{R}_{N_{\rm cap}}^{-} = R_{N_{\rm cap}}^{-}.
\end{array}
\label{eq.3.2.1.7}
\end{equation}

\subsection{Amplitudes, coefficients of penetrability and reflection concerning to barrier, test
\label{sec.3.2.2}}

We can also find the summed amplitude $A_{R, {\rm ext}}$ of all waves reflected from the external barrier region
(from the external turning point $r_{\rm tp,ext}$ to $r_{N-1}$)
and propagated outside (which can characterize a potential scattering) as
\begin{equation}
\begin{array}{lll}
  A_{R, {\rm ext}} = \tilde{R}_{N-1}^{-}, &
  {\rm where}\;
  \tilde{R}_{N_{\rm tp,ext}}^{-} = R_{N_{\rm tp,ext}}^{-}
\end{array}
\label{eq.3.2.2.1}
\end{equation}
and the summed amplitude $A_{R, {\rm tun}}$ of all waves which are reflected just inside the potential region from $r_{\rm cap}$ to the external turning point $r_{\rm tp,ext}$
[i.e. they are propagated through the external barrier region (without any reflection), tunnel under the barrier,
may propagate up to the boundary at point $r_{\rm cap}$
and then are reflected back from this boundary]
as
\begin{equation}
\begin{array}{lll}
  A_{R, {\rm tun}} = A_{R, {\rm bar}} - A_{R, {\rm ps}}.
\end{array}
\label{eq.3.2.2.2}
\end{equation}
%
We estimate the amplitude of oscillations $A_{\rm osc}$ near capture with number $N_{\rm cap}$ as
\begin{equation}
\begin{array}{lllll}
  A_{\rm osc} (N_{\rm cap}) = \displaystyle\frac{1}{1 - \tilde{R}_{N_{\rm cap}-1}^{-} \tilde{R}_{N_{\rm cap}}^{+}}.
\end{array}
\label{eq.3.2.2.3}
\end{equation}

We define coefficients of penetrability $T_{\rm bar}$ and reflection $R_{\rm bar}$ concerning to the whole barrier
(i.e. the potential region from $r_{\rm cap}$ to $r_{N-1}$),
and add also definitions for
coefficient $R_{\rm ext}$ of reflection from the external part of the barrier
(i.e. region from $r_{\rm tp, ext}$ to $r_{N-1}$),
coefficient $R_{\rm tun}$ of reflection from the barrier region (i.e. region from $r_{\rm cap}$ to $r_{\rm tp, ext}$)
as
\begin{equation}
\begin{array}{lllll}
\vspace{0.7mm}
   T_{\rm bar} = \displaystyle\frac{k_{\rm cap}}{k_{N}}\; \bigl\|A_{T, {\rm bar}} \bigr\|^{2}, &
   R_{\rm bar} = \bigl\|A_{R, {\rm bar}}\bigr\|^{2}, \\
   R_{\rm ext} = \bigl\|A_{R, {\rm ext}}\bigr\|^{2}, &
   R_{\rm tun} =  \bigl\| A_{R, {\rm tun}}\bigr\|^{2}, &
   R_{\rm tun} = \bigl\| A_{R, {\rm tun}} \bigr\|^{2}.
\end{array}
\label{eq.3.2.3.1}
\end{equation}
%
We check the property
\begin{equation}
  T_{\rm bar} + R_{\rm bar} = 1,
\label{eq.3.2.3.2}
\end{equation}
%
%
which is used as test to indicate whether the MIR method gives the proper solution for the wave function.



%
Let us calculate sums of the amplitudes $\alpha_{j}^{(i)}$ and $\beta_{j}^{(i)}$.
For the sums of amplitudes for the region with number $j$, we get
\begin{equation}
\begin{array}{ll}
  \beta_{j} \equiv
  \displaystyle\sum\limits_{i=1} \beta_{j}^{(i)} =
  \tilde{T}_{j}^{-}\, \Bigl( 1 + \displaystyle\sum\limits_{i=1} (\tilde{R}_{j-1} \tilde{R}_{j}^{+})^{i} \Bigr) =
  \displaystyle\frac{\tilde{T}_{j}^{-}}{1 - \tilde{R}_{j-1} \tilde{R}_{j}^{+}}, \\

  \alpha_{j} \equiv
  \displaystyle\sum\limits_{i=1} \alpha_{j}^{(i)} =
  \tilde{R}_{j-1}\, \displaystyle\sum\limits_{i=1} \beta_{j}^{(i)} =
  \displaystyle\frac{\tilde{R}_{j-1} \tilde{T}_{j}^{-}}{1 - \tilde{R}_{j-1} \tilde{R}_{1}^{+}}.
\end{array}
\label{eq.3.2.4.1}
\end{equation}

\subsection{Probability of existence of the compound nucleus
\label{sec.3.2.5}}

%
To determine the probability of the existence of a compound nucleus, we calculate the integral of the square of the modulus of the wave function over the inner region to the barrier.
We define this region as the region between two internal turning points $r_{\rm int,1}$ and $r_{\rm int,2}$.
In the region of above-barrier energies, we have:
\begin{equation}
\begin{array}{ll}
  & \displaystyle\int\limits_{r_{\rm int,1}}^{r_{\rm int,2}} \|\chi(r)\|^{2} dr =

  \displaystyle\sum\limits_{j=1}^{n_{\rm int}}
  \displaystyle\int\limits_{r_{j-1}}^{r_{j}}
    \Bigl\|
      \displaystyle\sum\limits_{i=1} \alpha_{j}^{(i)} e^{ik_{j}r} +
      \displaystyle\sum\limits_{i=1} \beta_{j}^{(i)} e^{-ik_{j}r} \Bigr\|^{2} dr = \\

  = &
  \displaystyle\sum\limits_{j=1}^{n_{\rm int}}
  \Bigl\{
    \bigl( \|\alpha_{j}\|^{2} + \|\beta_{j}\|^{2} \bigr)\, \Delta r +
    \displaystyle\frac{\|\alpha_{j}\beta_{j}\|} {k_{j}}\,
      \sin(\theta_{\alpha_{j}} - \theta_{\beta_{j}} + 2k_{j}r)
    \Bigr\|_{r_{j-1}}^{r_{j}}
  \Bigr\},
\end{array}
\label{eq.3.2.5.1}
\end{equation}
where $\theta_{\alpha_{j}}$ and $\theta_{\beta_{j}}$ are phases of the amplitudes $\alpha_{j}$ and $\beta_{j}$, respectively.
If the spatial region includes tunneling, then the formulas above should be rewritten in the complex form
(where $k_{j}$ are complex numbers):
%
%
\begin{equation}
\begin{array}{lllll}
  & \displaystyle\int\limits_{0}^{r_{\rm int,max}} \|\chi(r)\|^{2}\; dr =

  \displaystyle\sum\limits_{j=1}^{n_{\rm int,max}}
  \displaystyle\int\limits_{r_{j-1}}^{r_{j}}
    \bigl\| \alpha_{j} e^{ik_{j}r} + \beta_{j} e^{-ik_{j}r} \bigr\|^{2} \;dr = \\

%

  = &
  \displaystyle\sum\limits_{j=1}^{n_{\rm int,max}}
  \Bigl\{
    \bigl( \|\alpha_{j}\|^{2} + \|\beta_{j}\|^{2} \bigr)\, \Delta r +
    \displaystyle\frac{\alpha_{j}\beta_{j}^{*}} {2ik_{j}}\,  e^{2ik_{j}r}
      \Bigr\|_{r_{j-1}}^{r_{j}} -
    \displaystyle\frac{\alpha_{j}^{*}\beta_{j}} {2ik_{j}}\,  e^{-2ik_{j}r}
      \Bigr\|_{r_{j-1}}^{r_{j}}
  \Bigr\}.
\end{array}
\label{eq.3.2.5.2}
\end{equation}
%
%
In this paper we define the probability of the existence of the compound nucleus through the integral (\ref{eq.3.2.5.2}) over the spatial region between two internal turning points
(where the larger internal turning point is determined as the turning point of the barrier for sub-barrier energies, or as
the coordinate of the barrier maximum for above-barrier energies):
\begin{equation}
\begin{array}{llllll}
  P_{\rm cn} & \equiv &
  \displaystyle\int\limits_{r_{\rm int,1}}^{r_{\rm int,2}} \|\chi(r)\|^{2}\; dr = 

  \displaystyle\sum\limits_{j=1}^{n_{\rm int}}
  \Bigl\{
    \bigl( \|\alpha_{j}\|^{2} + \|\beta_{j}\|^{2} \bigr)\, \Delta r +
    \displaystyle\frac{\alpha_{j}\beta_{j}^{*}} {2ik_{j}}\,  e^{2ik_{j}r}
      \Bigr\|_{r_{j-1}}^{r_{j}} -
    \displaystyle\frac{\alpha_{j}^{*}\beta_{j}} {2ik_{j}}\,  e^{-2ik_{j}r}
      \Bigr\|_{r_{j-1}}^{r_{j}}
  \Bigr\}.
\end{array}
\label{eq.3.2.5.3}
\end{equation}
In case of the simplest barrier defined by Eqs.~(1) in Ref.~\cite{Maydanyuk_Zhang_Zou.2017.PRC} (and studied in Sect.~II in that paper)
one can write down $P_{\rm cn} (E)$ as
[see Eqs.~(6), (7), Ref.~\cite{Maydanyuk_Zhang_Zou.2017.PRC}]
%
\begin{equation}
\begin{array}{llllll}
  P_{\rm cn}^{\rm (without\, fusion)}  =
    P_{\rm osc}\, T_{\rm bar}\, P_{\rm loc}, & \\

  P_{\rm osc} = \|A_{\rm osc}\|^{2} =
  \displaystyle\frac{(k + k_{1})^{2}}
    {2k^{2} (1 -\cos (2k_{1}r_{1})) + 2k_{1}^{2}\,(1 + \cos (2k_{1}r_{1})) }, & \\

  T_{\rm bar} \equiv \displaystyle\frac{k_{1}}{k_{2}}\; \bigl\| T_{1}^{-} \bigr\|^{2}, & \\

  P_{\rm loc} = 2\, \displaystyle\frac{k_{2}}{k_{1}}\; \Bigl( r_{1} - \displaystyle\frac{\sin(2k_{1}r_{1})}{2k_{1}} \Bigr).
\end{array}
\label{eq.4.1.6}
\end{equation}
%
%
%
Comparing formula (\ref{eq.3.2.5.3}) with Eq.~(\ref{eq.4.1.6}),
%
we see that Eq.~(\ref{eq.3.2.5.3}) is much more complicated.
It is not possible to represent it exactly as a simple product of three coefficients of penetrability, oscillations and localization shown in Eqs.~\ref{eq.4.1.6}
(that corresponds to idea of Gamow for inverse processes, i.e. decays, as just direct multiplication of factor of oscillations and penetrability).
We also add formula for fast fusion for the simplest barrier as
\begin{equation}
\begin{array}{lllll}
  P_{\rm cn}^{\rm (fast\, fusion)} =

  \Bigl\| \displaystyle\sum\limits_{i=1} \beta_{1}^{(i)} \Bigr\|^{2}
  \displaystyle\int\limits_{0}^{r_{1}}
    \Bigl\| R_{0} e^{ik_{1}r} + e^{-ik_{1}r} \Bigr\|^{2}\, dr =



  \bigl\| T_{1}^{-} \bigr\|^{2}\, r_{1} =
  \displaystyle\frac{k_{2}\, r_{1}}{k_{1}} T_{\rm bar}.
\end{array}
\label{eq.3.1.7.6}
\end{equation}

\subsection{Cross-section of fusion and coefficients of fusion
\label{sec.3.2.6}}

%
In the description of capture of nuclei by nuclei,
there is a standard definition of the fusion cross section $\sigma$ based on the barrier penetrability $T_{\rm bar, l}$ and fusion probabilities $P_{l}$
(for example, see Ref.~\cite{Eberhard.1979.PRL},
where only $P_{l}=1$ or $P_{0}=0$
according to the sharp angular momentum cutoff in that paper,
see also Eq.~(3), (4) in Ref.~\cite{Maydanyuk.2015.NPA} for details):
\begin{equation}
\begin{array}{lll}
  \sigma_{\rm fus} (E) = \displaystyle\sum\limits_{l=0}^{+\infty} \sigma_{l}(E), &
  \sigma_{l}(E) = \displaystyle\frac{\pi\hbar^{2}}{2mE}\, (2l+1)\, T_{{\rm bar,} l}(E)\, P_{l},
\end{array}
\label{eq.3.2.6.1}
\end{equation}
%
%
where $\sigma_{l}$ is the partial cross-section of fusion at $l$,
$E$ is the energy of the relative motion of the incident nucleus relative to another nucleus.
To study the compound nucleus, we introduce a new definition of the partial fusion cross section in terms of the probability of the existence of a compound nucleus (\ref{eq.3.2.5.3}) as
\begin{equation}
  \sigma_{l} = \displaystyle\frac{\pi\hbar^{2}}{2mE}\, (2l+1)\, f_{l}(E)\, P_{\rm cn} (E),
\label{eq.3.2.6.2}
\end{equation}
%
%
where $f_{l}(E)$ is an additional factor that is needed to connect probability $P_{\rm cn} (E)$, penetrability $T_{{\rm bar,} l}(E)$ and the old factor of fusion $P_{l}$.
To find the explicit form of this coefficient, we consider a case of complete fusion, described by the old formula.
A similar result should give a coefficient at coefficients of fusion equal to one.
We obtain:
\begin{equation}
  f(E) = \displaystyle\frac{k_{\rm cap}}{k_{N}\, \|r_{\rm cap} - r_{\rm tp,in, 1}\|}.
\label{eq.3.2.6.3}
\end{equation}
%
%
Now, to study formation of the compound nucleus with slow fusion (i.e., without instantaneous fusion),
we vary these fusion coefficients in the spatial region between the capture point $r_{\rm cap}$ and the internal second turning point $r_{\rm int, 2}$.

\section{Analysis
\label{sec.analysis}}

We will choose reactions $\isotope[X]{C} + \isotope[X]{C} = \isotope[2X]{Mg}$~\cite{Gasques.2005.PRC} ($X = 10, 12, 14, 18, 20, 22, 24$) for the analysis.
The first indications on the possibility to synthesize more heavy elements from Carbon isotopes can be found in researches of
Hamada and Salpeter,
who estimated that \isotope[12]{C} would be converted to \isotope[24]{Mg} via pycnonuclear reactions above
density of $6 \cdot 10^{9}$ ${\rm g} \cdot {\rm cm}^{-3}$ \cite{Hamada_Salpeter.1961.AstrJ}.
That study was based on estimations of pycnonuclear reaction rates calculated by Cameron~\cite{Cameron.1959b.AstrJ}. 
Then estimates of densities of stellar medium for those reactions were improved
(for example, see estimates of Salpeter and Van Horn~\cite{Salpeter_VanHorn.1969.AstrJ}).
However, the densities quoted here are still quite uncertain.
Besides difficulty of an accurate calculation, finite temperatures and crystal imperfections can increase the rates significantly
(the critical density for carbon can be about $5 \cdot 10^{10}$ ${\rm g} \cdot {\rm cm}^{-3}$).
In our research we will focus on understanding of new quantum phenomena which exist in pycnonclear reactions and have not been studied yet by other researchers.
As inclusion of such effects can significantly change the rates of reactions and even the picture of participating mechanisms,
for brevity of calculations we will choose the density given by Hamada and Salpeter above.

\subsection{Potential of interaction between nuclei in lattice sites
\label{sec.analysis.3}}

\begin{figure}[htbp]
\centerline{\includegraphics[width=88mm]{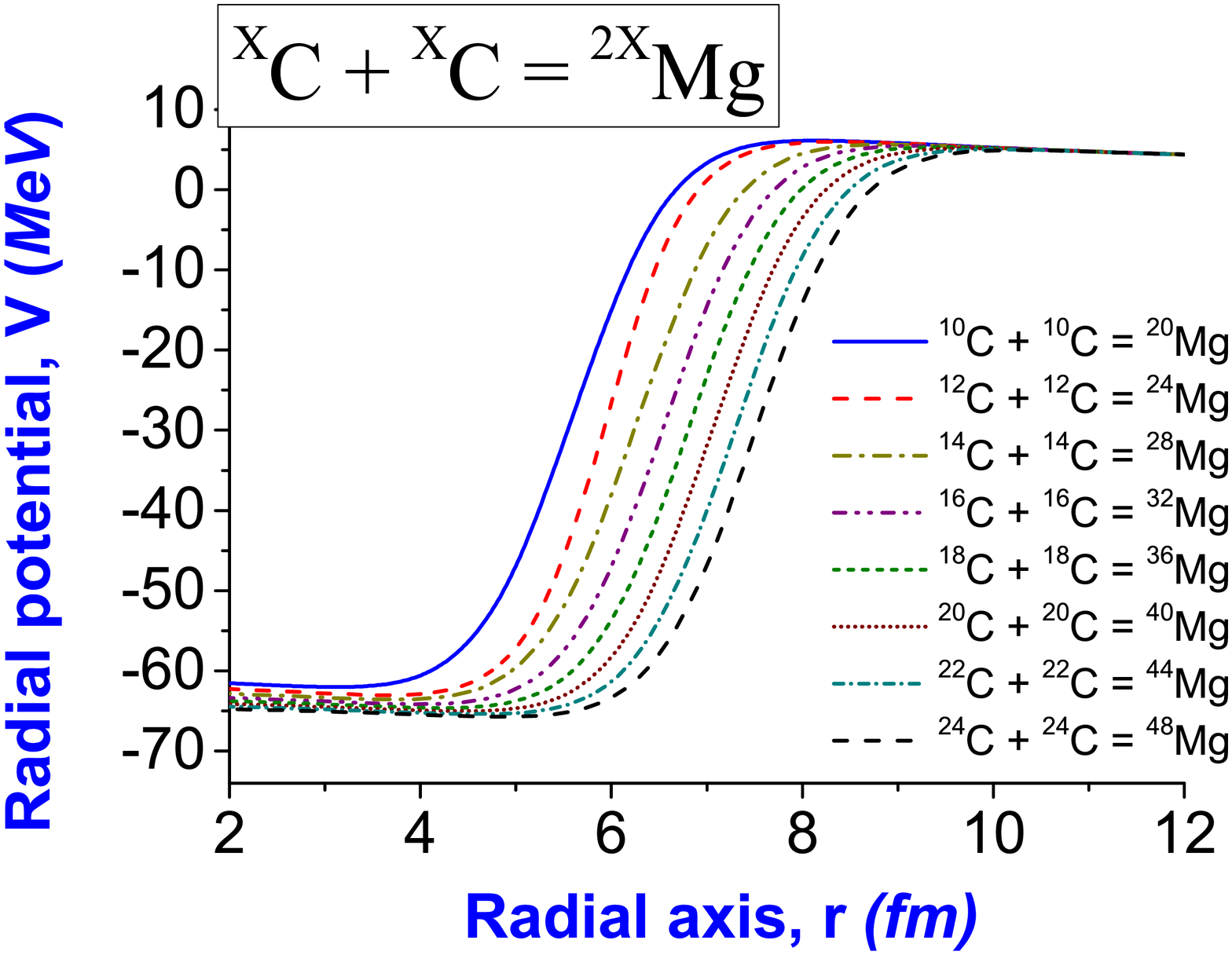}}
\caption{\small (Color online)
Potentials of interaction between two nuclei of carbon \isotope[X]{C}
[potentials and parameters are defined in Eq.~(\ref{eq.analysis.3.1})--(\ref{eq.analysis.3.3})].
\label{fig.3.1}}
\end{figure}

We define the potential of interactions between nuclei of carbon \isotope[X]{C} as
\begin{equation}
  V (r) = v_{c} (r) + v_{N} (r) + v_{l=0} (r),
\label{eq.analysis.3.1}
\end{equation}
where $v_{c} (r)$, $v_{N} (r)$ and $v_{l} (r)$ are Coulomb, nuclear and centrifugal components
have the form
%
\begin{equation}
\begin{array}{lll}
  \vspace{1mm}
  v_{N} (r) = - \displaystyle\frac{V_{R}} {1 + \exp{\displaystyle\frac{r-R_{R}} {a_{R}}}},
  \hspace{2mm}
  v_{l} (r) = \displaystyle\frac{l\,(l+1)} {2mr^{2}}, \\


  v_{c} (r) =
  \left\{
  \begin{array}{ll}
    \displaystyle\frac{Z_{1} Z_{2}\, e^{2}} {r}, &
      \mbox{at  } r \ge R_{c}, \\
    \displaystyle\frac{Z_{1} Z_{2}\, e^{2}} {2 R_{c}}\;
      \biggl\{ 3 -  \displaystyle\frac{r^{2}}{R_{c}^{2}} \biggr\}, &
      \mbox{at  } r < R_{c}.
  \end{array}
  \right.
\end{array}
\label{eq.analysis.3.2}
\end{equation}
Here, $V_{R}$ is strength of nuclear components defined in MeV as
%
\begin{equation}
\begin{array}{ll}
  V_{R} = -75.0\; \mbox{\rm MeV},
\end{array}
\label{eq.analysis.3.3}
\end{equation}
%
$R_{c}$ and $R_{R}$ are Coulomb and nuclear radiuses of nuclear system, $a_{R}$ is diffusion parameter,
$m$ is reduced mass defined in Eq.~(\ref{eq.analysis.4.5}).
We define parameters as
\cite{Maydanyuk_Shaulskyi.2022.EPJA,Maydanyuk.2010.PRC}
%
\begin{equation}
\begin{array}{llll}
  R_{R} = r_{R}\, (A_{1}^{1/3} + A_{2}^{1/3}), &
  R_{c} = r_{c}\, (A_{1}^{1/3} + A_{2}^{1/3}), &
  a_{R} = 0.44\; {\rm fm}, \\
  r_{R} = 1.30\;{\rm fm}, &
  r_{c} = 1.30\;{\rm fm}.
\end{array}
\label{eq.analysis.3.4}
\end{equation}
%
These potentials for isotopes of Carbon are shown in Fig.~\ref{fig.3.1}.
%
One can see a shape of internal well of the potential
(this internal well is absent in potentials used in Ref.~\cite{ShapiroTeukolsky.2004.book}, for example).
For brevity, we include maximums of barriers and minimums of wells for potentials of interaction between nuclei to Tabl.~~\ref{table.analysis.3.1}.
%
\begin{table}
\begin{center}
\begin{tabular}{|c|c|c|c|c|c|c|c|c|c|c|} \hline
  reaction $\isotope[X]{C} + \isotope[X]{C}$ &
    $r_{\rm min}$, fm & $V_{\rm min}$, MeV & $r_{\rm max}$, fm & $V_{\rm max}$, MeV & $R_{0}$, fm & $n_{A}$, $10^{-7}\; {\rm fm}^{-3}$ \\ \hline

  $\isotope[10]{C} + \isotope[10]{C}$ &
    3.36 & $-62.157$ & 7.98 & $+6.249$ & 87.06 & 3.617\, 027\, 31 \\

  $\isotope[12]{C} + \isotope[12]{C}$ &
    3.64 & $-63.018$ & 8.33 & $+5.972$ & 92.52 & 3.014\, 189\, 41 \\

  $\isotope[14]{C} + \isotope[14]{C}$ &
    3.92 & $-63.702$ & 8.68 & $+5.743$ & 97.40 & 2.583\, 590\, 92 \\

  $\isotope[16]{C} + \isotope[16]{C}$ &
    4.20 & $-64.258$ & 8.96 & $+5.552$ & 101.83 & 2.260\, 642\, 06 \\

  $\isotope[18]{C} + \isotope[18]{C}$ &
    4.48 & $-64.726$ & 9.24 & $+5.386$ & 105.91 & 2.009\, 459\, 61 \\

  $\isotope[20]{C} + \isotope[20]{C}$ &
    4.62 & $-65.133$ & 9.52 & $+5.242$ & 109.69 & 1.808\, 513\, 65 \\

  $\isotope[22]{C} + \isotope[22]{C}$ &
    4.90 & $-65.483$ & 9.80 & $+5.115$ & 113.23 & 1.644\, 103\, 31 \\

  $\isotope[24]{C} + \isotope[24]{C}$ &
    5.04 & $-65.792$ & 10.08 & $+5.001$ & 116.57 & 1.507\, 094\, 70 \\ \hline
\end{tabular}
\end{center}
\caption{%
Minimums of wells and maximums of barriers of potentials of interactions between two isotopes of carbon,
also distance $R_{0}$ between nuclei and their concentration $n_{A}$
[isotopes of carbon are chosen in accordance with Ref.~\cite{Afanasjev.2010.ADNDT}
on systematic study of astrophysical $S$-factors in fusion reactions for
\isotope{C}, \isotope{O}, \isotope{Ne}, \isotope{Mg},
parameters are determined for density
$\rho_{0} = 6 \cdot 10^{9}\, \displaystyle\frac{\mbox{g}} {\mbox{\rm cm}^{3}}$%
].
%
%
}
\label{table.analysis.3.1}
\end{table}
%

\subsection{Distance between nuclei in lattice sites
\label{sec.analysis.2}}

%
Let us determine at what distance the nuclei are in the stellar matter, taking the density of the matter into account.
We denote the distance between two close nuclei fixed in the lattice as $2\, R_{0}$, and put a ``scattering'' nucleus between these nuclei
(in analogy to Ref.~\cite{ShapiroTeukolsky.2004.book}, p.~90, Fig.~3.5).
The density $\rho_{0}$ in the sphere surrounding one nucleus of the lattice can be found as
the ratio of mass $m_{A}$ of the nucleus to the volume $V_{A}$ inside this sphere as
\begin{equation}
\begin{array}{llllll}
    \rho_{0} =
    \displaystyle\frac{m_{A}}{V_{A}} =
    \displaystyle\frac{A\, m_{u}}{4/3\, \pi\, R_{0}^{3}}

\end{array}
\label{eq.analysis.2.2}
\end{equation}
or
%
\begin{equation}
\begin{array}{llllll}
  R_{0} = \Bigl( \displaystyle\frac{A\, m_{u}}{4/3\, \pi\, \rho_{0}} \Bigr)^{1/3},
\end{array}
\label{eq.analysis.2.3}
\end{equation}
%
%
where $m_{u}$ is mass of nucleon,
$A$ is mass number of nucleus.
Concentration of nuclei $n_{A}$ is derived as
\begin{equation}
\begin{array}{llllll}
  n_{A} = \displaystyle\frac{\rho_{0}}{A\, m_{u}}.
\end{array}
\label{eq.analysis.2.6}
\end{equation}
For analysis of the pycnonuclear reactions $\isotope[X]{C} + \isotope[X]{C} = \isotope[2X]{Mg}$ we choose density,
estimated in Ref.~\cite{ShapiroTeukolsky.2004.book} 
\begin{equation}
\begin{array}{llllll}
  \rho_{0} = 6 \cdot 10^{9}\, \displaystyle\frac{\mbox{g}} {\mbox{\rm cm}^{3}}.
\end{array}
\label{eq.analysis.2.4}
\end{equation}
The derived distance $R_{0}$ and concentration $n_{A}$ for different isotopes of carbon at such a density are given in Tabl.~\ref{table.analysis.3.1}.

\subsection{Energy of zero-point vibrations of nuclei in lattice sites
\label{sec.analysis.4}}

A nucleus located in a lattice site and located between two nuclei with adjacent sites can oscillate and has a discrete spectrum of energy from such oscillations.
The approach to determine the energy levels of such a spectrum was proposed by Zel'dovich and investigated by other researchers.
In this approach the energy of zero-point vibrations of nucleus in lattice site is calculated as
[see Eqs.~(3.7.19), (3.7.20), \cite{ShapiroTeukolsky.2004.book}]:
%
%
\begin{equation}
\begin{array}{llllll}
\vspace{1mm}
  E_{0}^{\rm (zero)}  =
  \displaystyle\frac{\hbar w}{2} =
  \displaystyle\frac{\hbar\, Ze}{\sqrt{m\, R_{0}^{3}}}, &
  \Delta E = \displaystyle\frac{2\, Z^{2}e^{2}}{R_{0}}, &
  E_{\rm full} =
  E_{0}^{\rm (zero)} + \Delta E.
\end{array}
\label{eq.analysis.4.1}
\end{equation}
Here, $ E_{0}^{\rm (zero)}$ is energy of the ground state of harmonic oscillator relative to the potential minimum of this oscillator,
$\Delta E$ is shift of oscillator relative to zero value of the potential of interaction between nuclei
(i.e., distance between minimum of oscillator and zero value of potential of the interaction),
$E_{\rm full}$ is energy value of the ground state in system relative to zero value of the potential.
For example, for reaction $\isotope[12]{C} + \isotope[12]{C} = \isotope[24]{Mg}$ we obtain
%
\begin{equation}
\begin{array}{llllll}
  E_{0}^{\rm (zero)} = 0.021 808 06\; \mbox{\rm MeV}, &
  \Delta E = 0.567 872 37\; \mbox{\rm MeV}, &
  E_{\rm full}^{\rm (zero\, mode)} = 0.589 680 43\; \mbox{\rm MeV}.
\end{array}
\label{eq.analysis.4.2}
\end{equation}
%
For brevity of analysis, we call such a state (and the corresponding energy) as \emph{state of zero mode} (or state of zero-point vibrations of nuclei).

However, harmonic oscillator has not only the ground state, but the full discrete energy spectrum,
which are calculated as
%
\begin{equation}
\begin{array}{llllll}
\vspace{1mm}
  E_{n}^{\rm (zero)}  =
  \bigl( 2n + 1 \bigr) \cdot \displaystyle\frac{\hbar w}{2} =
  \bigl( 2n + 1 \bigr) \cdot E_{n=0}^{\rm (zero)} =
  \bigl( 2n + 1 \bigr)\, \displaystyle\frac{\hbar\, Ze}{\sqrt{m\, R_{0}^{3}}}.
\end{array}
\label{eq.analysis.4.4}
\end{equation}

Energy spectrum can be written down via density of matter $\rho_{0}$, instead of distance $R_{0}$.
Using Eq.~(\ref{eq.analysis.2.3}) and formula for reduced mass
%
\begin{equation}
\begin{array}{llllll}
  R_{0} = \Bigl( \displaystyle\frac{A\, m_{u}}{4/3\, \pi\, \rho_{0}} \Bigr)^{1/3}, &
  m = m_{\rm p}\, \displaystyle\frac{A_{1}\, A_{2}}{A_{1} + A_{2}}
\end{array}
\label{eq.analysis.4.5}
\end{equation}
from Eq.~(\ref{eq.analysis.4.1}) we obtain
(let us consider case of the same nuclei in lattice: $A_{1} = A_{2}$, and $A = A_{1}$):
%
\begin{equation}
\begin{array}{llllll}
\vspace{1mm}
  E_{0}^{\rm (zero)}  =
  c_{1} \cdot \displaystyle\frac{Z}{A}\, \sqrt{\rho_{0}}, &
  c_{1} = \hbar\, e \sqrt{\displaystyle\frac{8 \pi} {3\, m_{u} m_{\rm p}}}, \\

  \Delta E =
  c_{2} \cdot Z^{2} \Bigl( \displaystyle\frac{\rho_{0}}{A} \Bigr)^{1/3}, &
  c_{2} = 2 e^{2} \Bigl( \displaystyle\frac{4\pi}{3 m_{u}} \Bigr)^{1/3}.
\end{array}
\label{eq.analysis.4.7}
\end{equation}
We find new interesting property for nuclei of type $2Z = A$:
%
\begin{equation}
\begin{array}{llllll}
  E_{0}^{\rm (zero)}  =
  \displaystyle\frac{c_{1}}{2}\, \sqrt{\rho_{0}}, &

  \Delta E =
  c_{2} \cdot Z^{2} \Bigl( \displaystyle\frac{\rho_{0}}{2Z} \Bigr)^{1/3}.
\end{array}
\label{eq.analysis.4.8}
\end{equation}
Thus, according to this property, the spectra $E_{n}^{\rm (zero)}$ are the same for nuclei
\isotope[8]{Be}, \isotope[10]{B}, \isotope[12]{C}, \isotope[14]{N}, \isotope[16]{O}, \isotope[18]{F},
\isotope[20]{Ne}, \isotope[22]{Na}, \isotope[24]{Mg}, \isotope[26]{Si}, etc..
Those depend only on the chosen density in the stellar medium.
In Tabl.~\ref{table.Zeldovich.zeromode} energy values are presented for the first 10 states of zero-point vibrations calculated by Eqs.~(\ref{eq.analysis.4.1})
for reactions $\isotope[X]{C} + \isotope[X]{C}$.
%
%
\begin{table}
\begin{center}
\begin{tabular}{|c|c|c|c|c|c|c|} \hline
  No. & Energy, $E_{n}^{\rm (zero)}$, MeV & Energy, $E_{\rm full}^{\rm (zero)}$, MeV \\ \hline
 1 & 0.021808061833736 & 0.589680437522993 \\
 2 & 0.065424185501208 & 0.633296561190465 \\
 3 & 0.109040309168680 & 0.676912684857937 \\
 4 & 0.152656432836153 & 0.720528808525410 \\
 5 & 0.196272556503626 & 0.764144932192882 \\
 6 & 0.239888680171098 & 0.807761055860354 \\
 7 & 0.283504803838570 & 0.851377179527827 \\
 8 & 0.327120927506043 & 0.894993303195299 \\
 9 & 0.370737051173515 & 0.938609426862772 \\
10 & 0.414353174840987 & 0.982225550530244 \\
  \hline
\end{tabular}
\end{center}
\caption{Energy levels for the first 10 states of zero-point vibtations calculated by Eqs.~(\ref{eq.analysis.4.1})
for reactions $\isotope[X]{C} + \isotope[X]{C}$
}
\label{table.Zeldovich.zeromode}
\end{table}

\subsection{Energy spectra of zero-point vibrations:
Method of Multiple internal reflections versus Zel'dovich's approach
\label{sec.analysis.zeropoints}}

Formalism of the method of Multiple Internal Reflections can be further investigated,
with aim to calculate spectrum of energy in the states of zero-point vibrations.
Radial wave function in the asymptotic region has the form as
%
\begin{equation}
\begin{array}{llllll}
  \chi (r) =  e^{-ikr} + A_{R}\, e^{+ikr}.
\end{array}
\label{eq.analysis.zeropoints.1}
\end{equation}
In particular, the wave function at point $R_{0}$ has the same form
(where the scattered nucleus is at the minimum of potential between two nuclei in the lattice sites).
However, discreteness of the spectrum of energy imposes an additional requirement that
the total wave function should be equal to zero (for odd states) or
take the maximum value in module (for even states) at that point%
We find
%
\begin{equation}
\begin{array}{llllll}
\vspace{1.5mm}
  1) &
  \chi (R_{0}) =  e^{-ikR_{0}} + A_{R}\, e^{+ikR_{0}} = e^{-ikR_{0}} + e^{+ikR_{0}}, &
  A_{R} = + 1, \\

  2) &
  \chi (R_{0}) =  e^{-ikR_{0}} + A_{R}\, e^{+ikR_{0}} = e^{-ikR_{0}} - e^{+ikR_{0}}, &
  A_{R} = - 1.
\end{array}
\label{eq.analysis.zeropoints.2}
\end{equation}

\begin{figure}[htbp]
\centerline{\includegraphics[width=88mm]{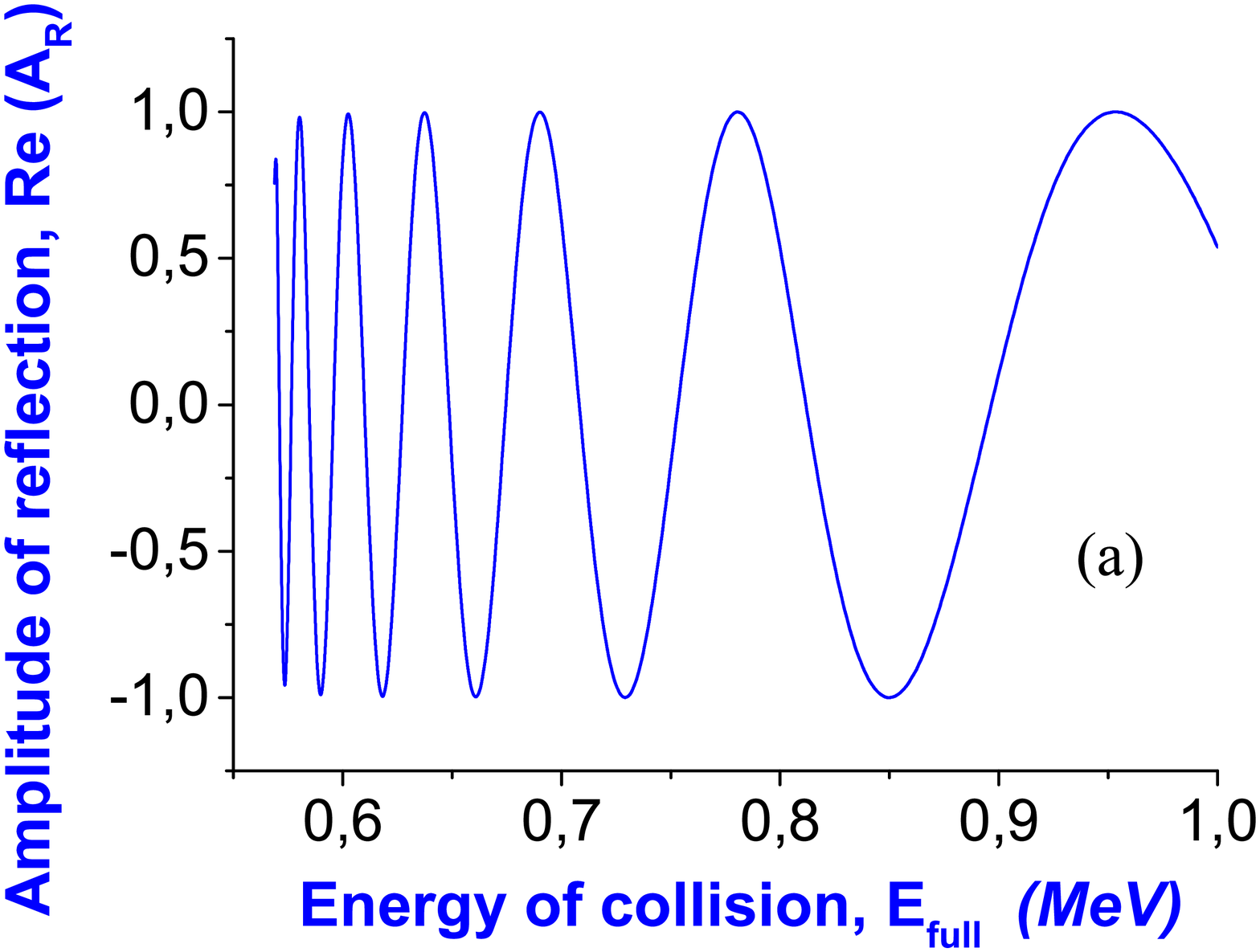}
\hspace{-1mm}\includegraphics[width=88mm]{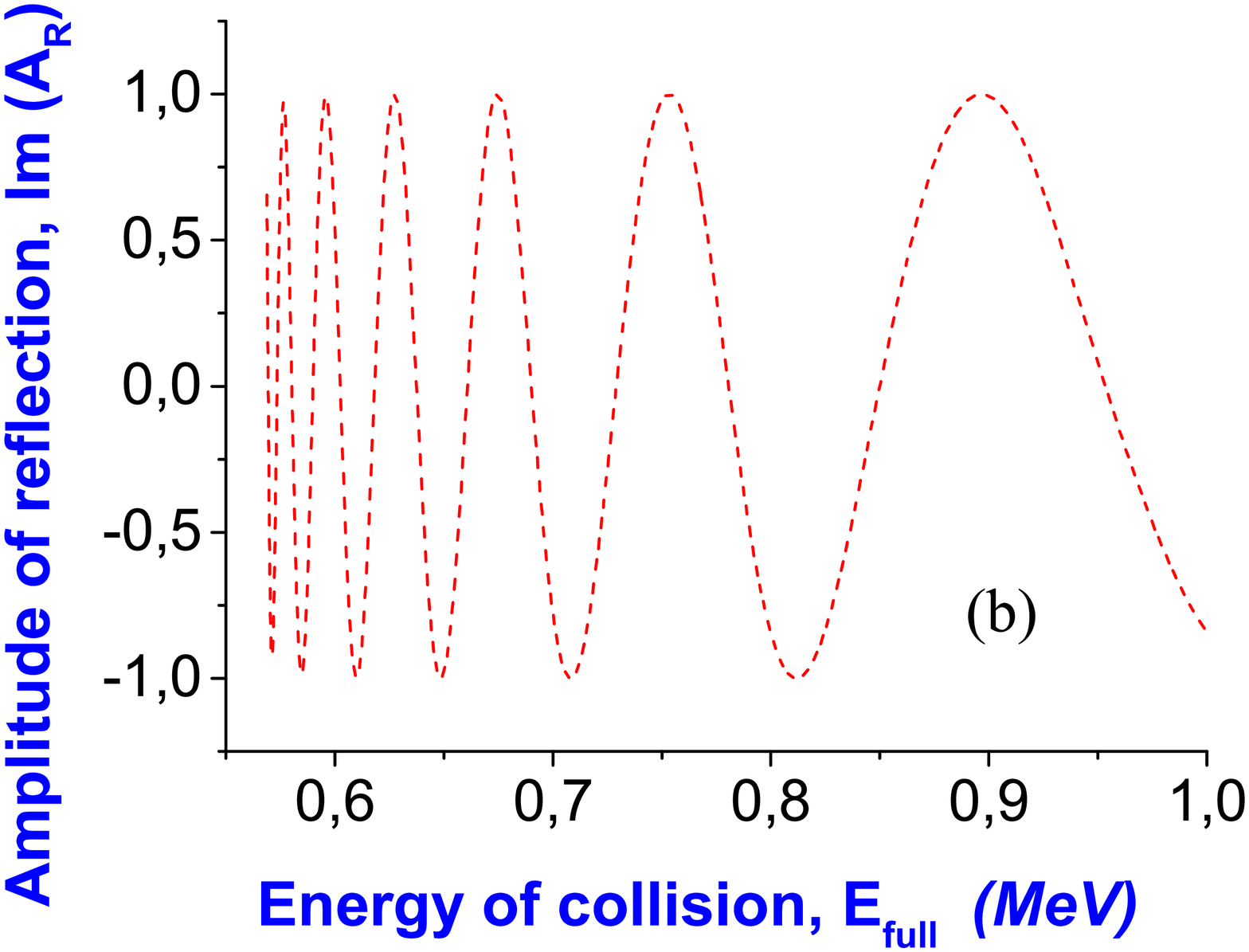}}
\caption{\small (Color online)
The real and imaginary parts of the amplitude of reflection, $A_{R} (r)$, calculated for
$\isotope[12]{C} + \isotope[12]{C}$
at point $R_{0}$ in the region of energy of relative motion $E_{\rm kin} = 0.55$ -- 1.0~MeV
[potential and parameters are defined  by Eqs.~(\ref{eq.analysis.3.1})--(\ref{eq.analysis.3.3}),
$R_{0}$ is determined by formula (\ref{eq.analysis.2.3}), its value is specified in Tabl.~\ref{table.analysis.3.1}].
%
\label{fig.9}}
\end{figure}

The amplitude of reflection calculated by this way at point $R_{0}$ for
$\isotope[12]{C} + \isotope[12]{C}$ is shown in Fig.~\ref{fig.9}.
%
One can see that each function has its own maxima and minima at close energy values.
According to Eq.~(\ref{eq.analysis.zeropoints.2}), energy levels can be determined from equality to zero of the imaginary part of such an amplitude [see Fig.~\ref{fig.9}~(b)].
%
\begin{equation}
\begin{array}{llllll}
\vspace{1.5mm}
  1) &
  \mbox{\rm even states:} &
  A_{R} = + 1, &
  \Re(A_{R}) = +1,  & \Im(A_{R}) = 0,\\

  2) &
  \mbox{\rm odd states:} &
  A_{R} = - 1, &
  \Re(A_{R}) = -1,  & \Im(A_{R}) = 0.
\end{array}
\label{eq.analysis.zeropoints.3}
\end{equation}

In Tabl.~\ref{table.5} the energy values calculated in the discrete energy spectrum for reactions with carbon isotopes
$\isotope[X]{C} + \isotope[X]{C} \to \isotope[2X]{Mg}$ are presented.
Parity of each state is determined by sign of the real part of amplitude of reflection.
%
\begin{table}
\begin{center}
\begin{tabular}{|c|c|c|c|c|c|c|l|l|l|l|} \hline
  No.    & \isotope[10]{C} + \isotope[10]{C}
         & \isotope[12]{C} + \isotope[12]{C}
         & \isotope[14]{C} + \isotope[14]{C}
         & \isotope[16]{C} + \isotope[16]{C} \\ \hline
 1  &  0.517434869739479  & 
     0.517434869739479 & 
     0.517434869739479 & 
     0.527054108216433 
\\

 2  &  0.536673346693387 & 
       0.536673346693387 & 
       0.536673346693387 & 
       0.536072144288577 
\\

 3  & 0.546292585170341 & 
      0.546292585170341 & 
      0.546292585170341 & 
      0.545090180360721 
\\

 4  & 0.565531062124249 & 
      0.555911823647295 & 
      0.555911823647295 & 
      0.554108216432866 
\\

 5  & 0.584769539078156 & 
      0.575150300601202 & 
      0.565531062124249 & 
      0.572144288577154 
\\

 6 & 0.613627254509018 & 
     0.594388777555110 & 
     0.575150300601202 & 
     0.581162324649299 
\\

 7 & 0.642484969939880 & 
     0.613627254509018 & 
     0.594388777555110 & 
     0.599198396793587 
\\

 8 & 0.680961923847695 & 
     0.642484969939880 & 
     0.623246492985972 & 
     0.626252505010020 
\\

 9  & 0.738677354709419 & 
      0.680961923847695 & 
      0.652104208416834 & 
      0.653306613226453 
\\

10  & 0.815631262525050 & 
      0.729058116232465 & 
      0.690581162324649 & 
      0.689378757515030 
\\

 11  & 0.950300601202405 & 
       0.806012024048096 & 
       0.729058116232465 & 
       0.734468937875752 
\\

 12 & 1.27735470941884 & 
     0.911823647294589 & 
     0.796392785571142 & 
     0.797595190380762 
\\

13 & 2.23927855711423 & 
     1.11382765531062 & 
     0.892585170340681 & 
     0.878757515030060 
\\

 14 & 3.69178356713427 & 
      2.76833667334669 & 
      1.04649298597194 & 
      1.02304609218437 
\\

15 & --- &
   4.08617234468938 & 
   1.64288577154309 & 
   1.39278557114228 
\\

16 & --- & --- &
  3.04729458917836 & 
  1.99699398797595 
\\

17 & --- & --- &
  4.28817635270541 & 
  3.20541082164329 
\\

18 & --- & --- & --- &
  4.37775551102204 
\\
\hline
\end{tabular}
\end{center}
\caption{%
Energy levels of zero-point vibrations $E_{\rm zero}^{\rm (mir)}$
for reactions $\isotope[X]{C} + \isotope[X]{C}$
(presented data are in MeV, below than 5~MeV)
calculated by the method of Multiple internal reflections (see Sec.~\ref{sec.analysis.zeropoints}, for details).
Distances between each two adjacent energy levels are radically different from the spectrum of harmonic oscillator,
obtained in the approach of Zel'dovich and his colleagues (see Tabl.~\ref{table.Zeldovich.zeromode}).
The amplitude $\Re (A_{R})$ for each level takes positive or negative value that determines even or odd state in the discrete energy spectrum.
The energy levels in table belong to the discrete spectrum for each studied system and
there are few energy levels below energy of the zero-point vibrations in the ground state
$E_{\rm full, 0}^{\rm (zero)}$ calculated in the approach of Zel'dovich and his colleagues
[see Eq.~(\ref{eq.analysis.4.1})].
Accuracy of determination of energy levels is estimated from $|\Re (A_{R})| \approx 1$ (it can be improved without limit).
Summation of $[\Re (A_{R})]^{2} + [\Im (A_{rm R})]^{2}$ is additional estimation of accuracy of the method MIR in determination of obtained digits of the amplitude.
%
}
\label{table.5}
\end{table}

%
\begin{table}
\begin{center}
\begin{tabular}{|c|c|c|c|c|c|c|l|l|l|l|} \hline
  No.    & \isotope[18]{C} + \isotope[18]{C}
         & \isotope[20]{C} + \isotope[20]{C}
         & \isotope[22]{C} + \isotope[22]{C}
         & \isotope[24]{C} + \isotope[24]{C} \\ \hline
 1 & 0.517434869739479 & 
     0.527054108216433 & 
     0.536673346693387 & 
     0.517434869739479 
\\

   2 & 0.527054108216433 & 
       0.546292585170341 & 
       0.546292585170341 & 
       0.527054108216433 
\\

  3 & 0.536673346693387 & 
      0.555911823647295 & 
      0.555911823647295 & 
      0.536673346693387 
\\

  4 & 0.546292585170341 & 
      0.565531062124249 & 
      0.565531062124249 & 
      0.584769539078156 
\\

  5 & 0.555911823647295 & 
      0.584769539078156 & 
      0.575150300601202 & 
      0.594388777555110 
\\

  6 & 0.565531062124249 & 
      0.604008016032064 & 
      0.594388777555110 & 
      0.613627254509018 
\\

  7 & 0.575150300601202 & 
      0.623246492985972 & 
      0.604008016032064 & 
      0.632865731462926 
\\

  8 & 0.584769539078156 & 
      0.642484969939880 & 
      0.623246492985972 & 
      0.661723446893788 
\\

  9 & 0.594388777555110 & 
      0.671342685370741 & 
      0.652104208416834 & 
      0.690581162324649 
\\

 10 & 0.613627254509018 & 
      0.709819639278557 & 
      0.680961923847695 & 
      0.729058116232465 
\\

 11 & 0.632865731462926 & 
      0.748296593186373 & 
      0.719438877755511 & 
      0.767535070140281 
\\

 12 & 0.661723446893788 & 
      0.806012024048096 & 
      0.757915831663327 & 
      0.825250501002004 
\\

 13 & 0.738677354709419 & 
      0.882965931863727 & 
      0.815631262525050 & 
      0.911823647294589 
\\

 14 & 0.796392785571142 & 
      1.00801603206413 & 
      0.892585170340681 & 
      1.02725450901804 
\\

 15 & 0.882965931863727 & 
      1.27735470941884 & 
      1.01763527054108 & 
      1.29659318637275 
\\

 16 & 1.00801603206413 & 
      2.24889779559118 & 
      1.27735470941884 & 
      2.20080160320641 
\\

 17 & 1.30621242484970 & 
      3.26853707414830 & 
      2.24889779559118 & 
      3.14348697394790 
\\

 18 & 2.18156312625251 & 
      4.29779559118237 & 
      3.22044088176353 & 
      4.08617234468938 
\\

 19 & 3.26853707414830 & 
      --- &
      4.21122244488978 & 
      ---
\\ \hline
\end{tabular}
\end{center}
\caption{(Continuation of Tabl.~\ref{table.5})}
\label{table.5b}
\end{table}

\subsection{Probabilities of formation of compound nuclei in pycnonuclear reactions
\label{sec.analysis.9}}

Our analysis has shown that the coefficients of penetrability and reflection increases monotonously with
the energy of the incident nucleus~\cite{Maydanyuk_Shaulskyi.2022.EPJA}.
There are no maximums and minimums in such dependencies.
This means that penetrability and reflection themselves cannot indicate on possible existence of some definite states of more heavy nuclei synthesized in pycnonuclear reactions in stars.
Such behaviour of these characteristics is in full agreement with previous studies on the capture of $\alpha$ particles by
nuclei~\cite{Maydanyuk.2015.NPA,Maydanyuk_Zhang_Zou.2017.PRC}.

\begin{figure}[htbp]
\centerline{\includegraphics[width=88mm]{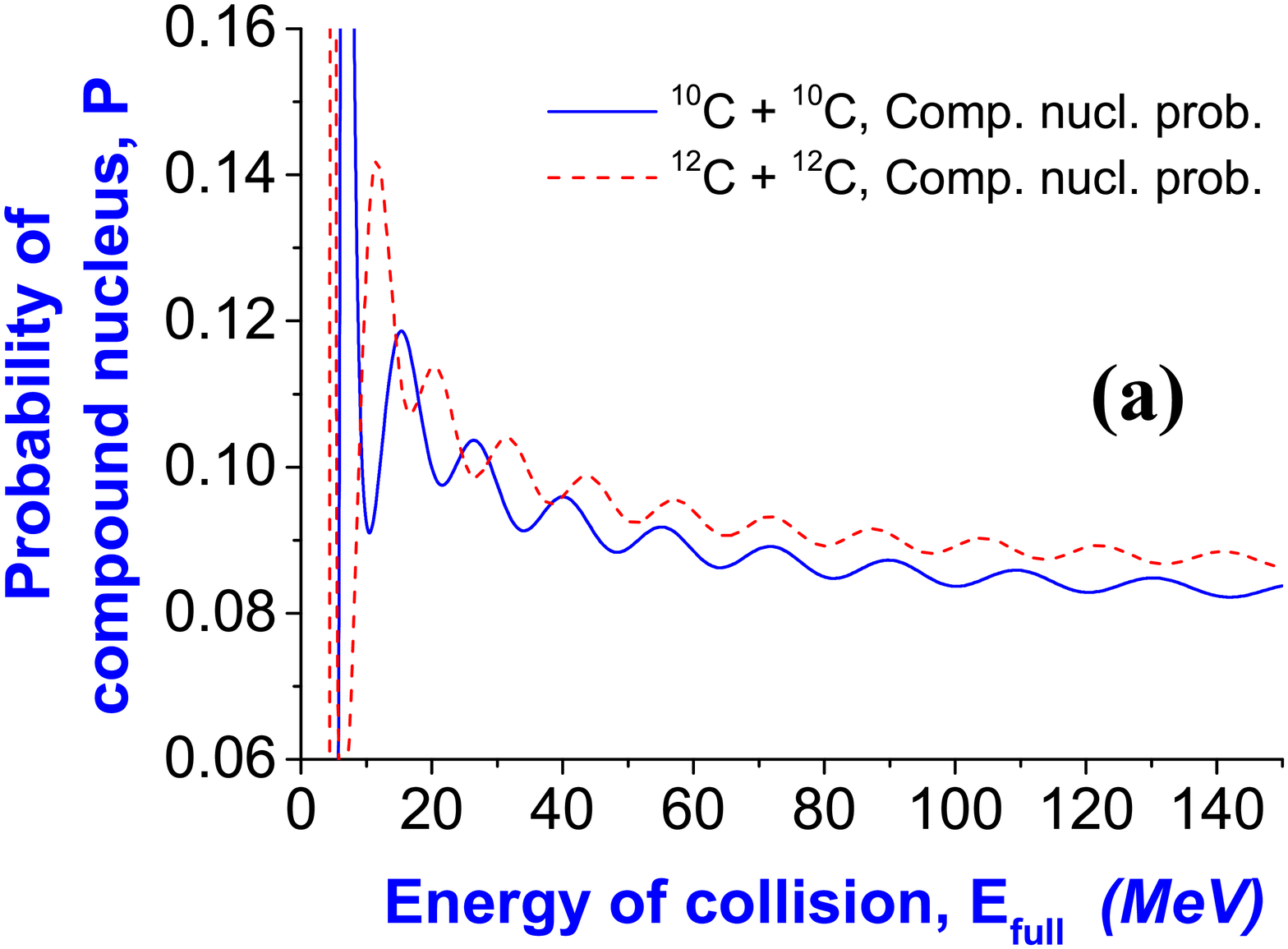}
\hspace{-3mm}\includegraphics[width=88mm]{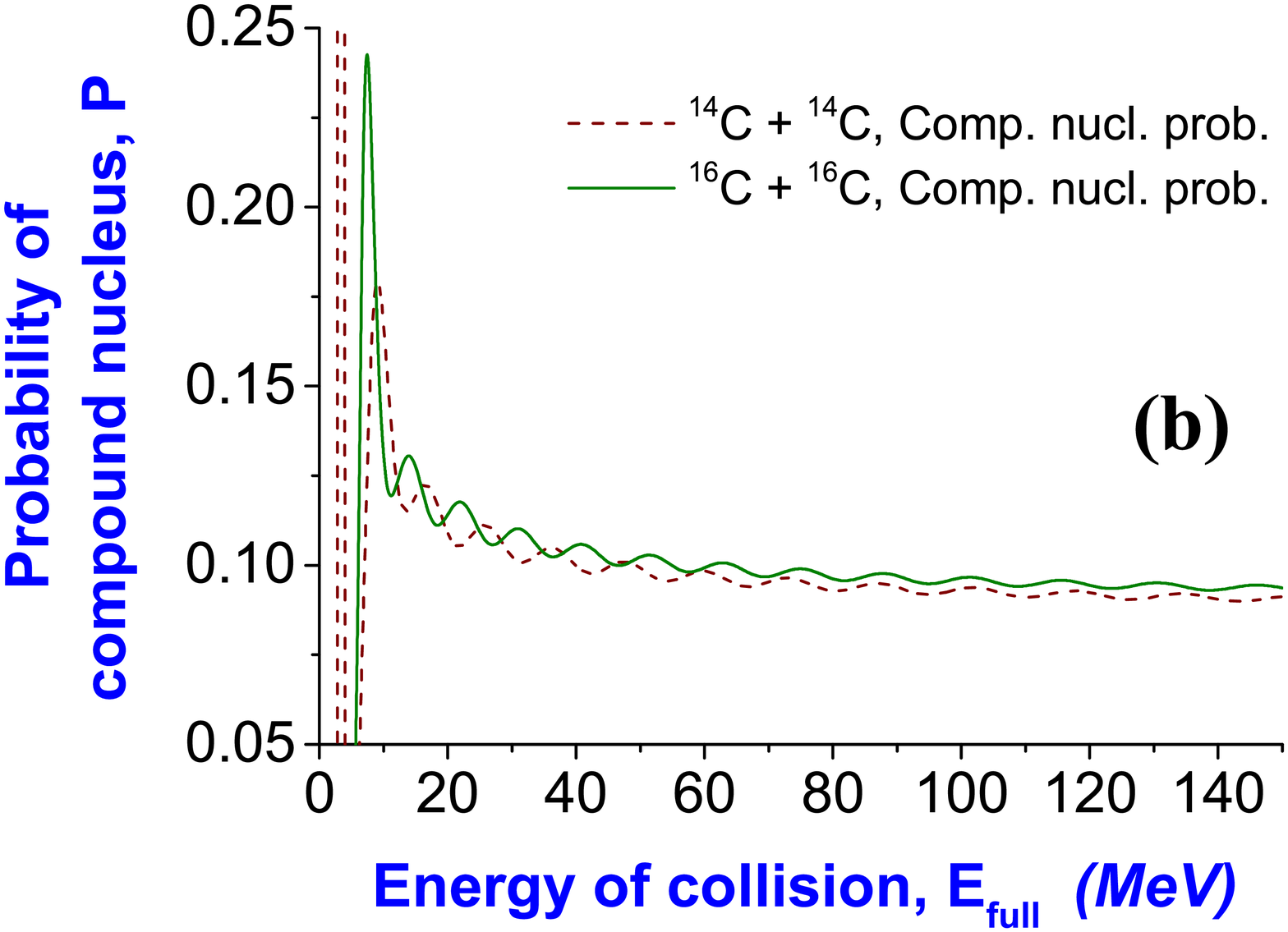}}
\centerline{\includegraphics[width=88mm]{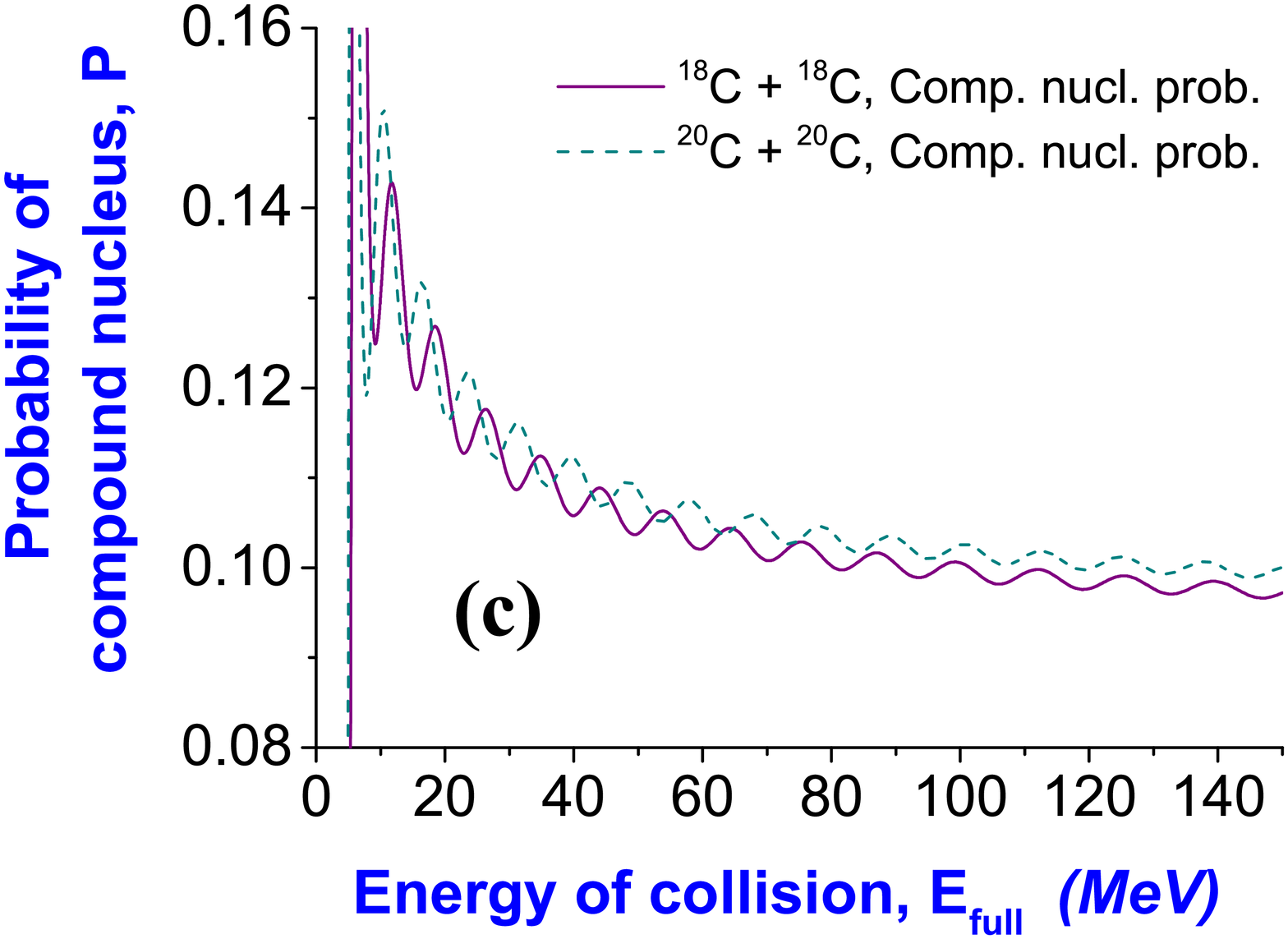}
\hspace{-3mm}\includegraphics[width=88mm]{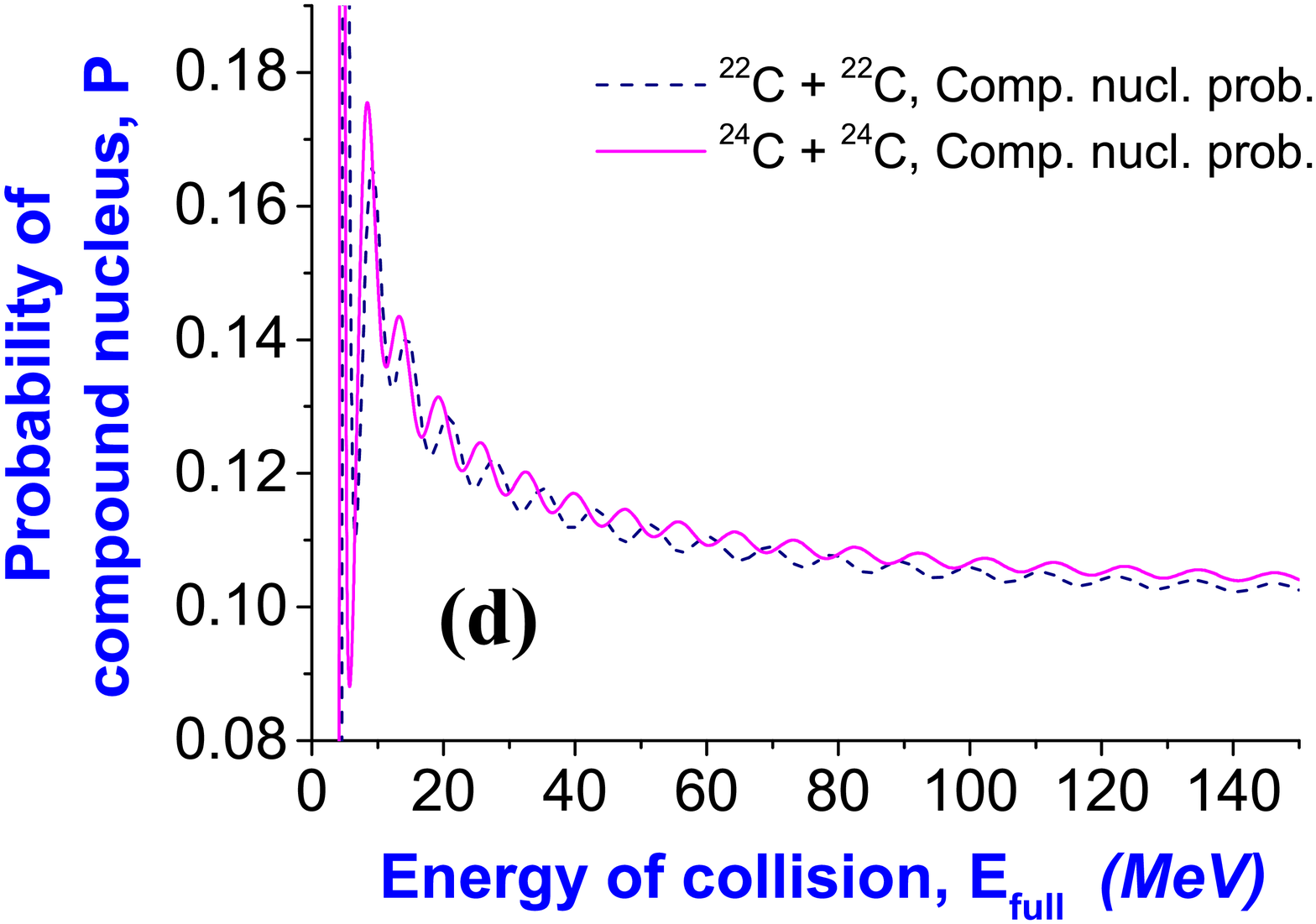}}
\caption{\small (Color online)
%
Probabilities of formation of compound nucleus $P_{\rm cn}$ in dependence on energy for incident isotopes of Carbon in reactions
$\isotope[10]{C} + \isotope[10]{C}$, $\isotope[12]{C} + \isotope[12]{C}$ (a),
$\isotope[14]{C} + \isotope[14]{C}$, $\isotope[16]{C} + \isotope[16]{C}$ (b),
$\isotope[18]{C} + \isotope[18]{C}$, $\isotope[20]{C} + \isotope[20]{C}$ (c) and
$\isotope[22]{C} + \isotope[22]{C}$, $\isotope[24]{C} + \isotope[24]{C}$ (d) in lattice
[potentials and parameters are defined in Eq.~(\ref{eq.analysis.3.1})--(\ref{eq.analysis.3.3})].
One can clearly see presence of maxima of such probabilities for all studied isotopes.
Energies corresponding to maxima of such probabilities are given in Tabl.~\ref{table.2}.
\label{fig.3.4}}
\end{figure}

Another important quantum characteristic is probability of formation of a compound nucleus which can be created during the studied reactions with nuclei.
In Fig.~\ref{fig.3.4} we present such probabilities for isotopes of Carbon calculated by our method.
%
%
In these figure one can clearly see presence of maxima in probabilities at certain definite energies for all studied isotopes of Carbon.
This means that at such energies compound nuclei are formed with maximum probability.
Another conclusion from such calculations is that at such energies the formation of the new \isotope[X]{Mg} nuclei is much more probable,
than at energies of zero mode vibrations determined by the theory of pycnonuclear reactions in compact stars.
This result gives an entirely new picture of pycnonuclear reactions in compact stars.
These maxima are explained by the need to take into account the further propagation of quantum fluxes in the potential region,
in contrast to the existing modern description of pychonuclear reactions, where these fluxes are interrupted at internal turning point and disappeared in nuclear region.
Quantum mechanics requires such a consideration, which strictly requires the continuity of the wave function in the full region of its definition
and the conservation of fluxes in the full region of definition of the wave function.
In  Tabl.~\ref{table.2} we present energies of the quasibound states for reactions with isotopes of Carbon up to 150~MeV.
\begin{table}
\hspace{-20mm}
\begin{tabular}{|c|c|c|c|c|c|c|c|c|c|} \hline
  No. & $\isotope[10]{C} + \isotope[10]{C}$ & $\isotope[12]{C} + \isotope[12]{C}$ &
  $\isotope[14]{C} + \isotope[14]{C}$ & $\isotope[16]{C} + \isotope[16]{C}$ &
  $\isotope[18]{C} + \isotope[18]{C}$ & $\isotope[20]{C} + \isotope[20]{C}$ &
  $\isotope[22]{C} + \isotope[22]{C}$ & $\isotope[24]{C} + \isotope[24]{C}$
  \\ \hline

  1   &   0.63471 &   4.88176 &   9.06212 &   7.27054 &   6.37475 &   5.47896 &   5.18036 &   4.58317 \\
  2   &  15.33267 &  11.45090 &  16.52705 &  13.83968 &  11.74950 &  10.55511 &   9.06212 &   8.46493 \\
  3   &  26.38076 &  20.40882 &  25.78357 &  21.90180 &  18.31864 &  16.52705 &  14.43687 &  13.24248 \\
  4   &  40.11623 &  31.45691 &  36.23447 &  30.85972 &  26.38076 &  23.69339 &  20.70741 &  19.21443 \\
  5   &  55.04609 &  43.69940 &  47.58116 &  40.71343 &  34.74148 &  31.15832 &  27.57515 &  25.48497 \\

  6   &  71.76754 &  57.13627 &  59.82365 &  51.46293 &  43.99800 &  39.51904 &  35.04008 &  32.35271 \\
  7   &  89.68337 &  71.76754 &  72.96192 &  62.80962 &  53.85170 &  48.47695 &  42.80361 &  39.81764 \\
  8   & 109.39078 &  87.29459 &  86.99599 &  74.75351 &  64.30261 &  57.73347 &  51.16433 &  47.58116 \\
  9   & 130.29259 & 104.01603 & 101.62725 &  87.59319 &  75.35070 &  67.88577 &  60.12224 &  55.64329 \\
  10  &     --    & 121.93186 & 117.45291 & 101.03006 &  86.99599 &  78.03808 &  69.37000 &  64.30261 \\

  11  &     --    &     --    & 134.17435 & 115.36273 &  98.93988 &  89.08617 &  79.23246 &  72.96192 \\
  12  &     --    &     --    &     --    & 130.29259 & 112.07816 & 100.43287 &  89.08617 &  82.51703 \\
  13  &     --    &     --    &     --    & 146.11824 & 125.21643 & 112.37675 &  99.83567 &  92.07214 \\
  14  &     --    &     --    &     --    &     --    & 139.25050 & 124.91784 & 110.88377 & 102.22445 \\
  15  &     --    &     --    &     --    &     --    &     --    & 137.75752 & 122.23046 & 112.67535 \\

  16  &     --    &     --    &     --    &     --    &     --    &     --    & 134.17435 & 123.42485 \\
  17  &     --    &     --    &     --    &     --    &     --    &     --    & 146.11824 & 134.77154 \\
  18  &     --    &     --    &     --    &     --    &     --    &     --    &     --    & 146.11824 \\
  \hline
\end{tabular}
\caption{Energies of the quasibound states of the compound nuclear systems formed in reactions with isotopes of Carbon
\isotope[10]{C}, \isotope[12]{C}, \isotope[14]{C}, \isotope[16]{C},
\isotope[18]{C}, \isotope[20]{C}, \isotope[22]{C}, \isotope[24]{C},
calculated by the method of multiple internal reflections up to 150~MeV
%
[for each calculation, we check test of $|T_{\rm bar} + R_{\rm bar}| = 1$ and obtain accuracy of $10^{-14}$ of its confirmation].
%
Comparing these energies with maximums of the potential barriers for all studied systems given in Tabl.~\ref{table.analysis.3.1},
we find that
only first quasibound energies for
$\isotope[10]{C} + \isotope[10]{C}$,
$\isotope[12]{C} + \isotope[12]{C}$,
$\isotope[24]{C} + \isotope[24]{C}$ are smaller than barrier maximums for these nuclear systems.
That means that at such energies
the compound nuclear systems have barriers which prevent decays going through tunneling phenomenon.
}
\label{table.2}
\end{table}
Only first quasibound energies for
$\isotope[10]{C} + \isotope[10]{C}$,
$\isotope[12]{C} + \isotope[12]{C}$,
$\isotope[24]{C} + \isotope[24]{C}$ are smaller than the barrier maximums for these nuclear systems.
This means that at such energies
compound nuclear systems are the most stable and are transformed to new synthesized isotopes of Magnesium \isotope[20]{Mg}, \isotope[24]{Mg} and \isotope[48]{Mg} with large probability.
There is a simple way to estimate half-lives of these obtained heavier nuclei using Gamow's approach
(well developed in the problem of nuclear decays) or the method of Multiple internal reflections for higher precision
(we omit these calculations in this paper).

\section{Plasma screening in nuclear reactions
\label{sec.screening}}

It is well known that nuclear reactions in compact stars,
which contain matter of high density, can be strongly modified by plasma physics effects~\cite{Salpeter_VanHorn.1969.AstrJ}.
So, a natural question appears in how much of the results presented above are changed after taking effects of plasma screening into account.
We will follow Ref.~\cite{Kravchuk.2014.PRC},
where effects of plasma screening in thermonuclear fusion reactions in dense nuclear matter in stars were studied.
Here, in addition of physical analysis, authors provided clear formalism for use and implementations to other researches.
So, we will estimate influence of plasma screening on pycnonuclear reactions on the basis of isotope \isotope[12]{C} and
we use that research as a basis for our analysis.

Following Ref.~\cite{Kravchuk.2014.PRC} [see Eq.~(7) in that paper],
we define Coulomb potential $U_{C}(r)$ for colliding nuclei in the standard form
\begin{equation}
\begin{array}{llllll}
  U_{C,\, full} (r) =
  v_{C} (r) + H(r),
\end{array}
\label{eq.screening.1.1}
\end{equation}
where
$v_{C}(r)$ is pure Coulomb potential without screening,
$H(r)$ is the mean-field plasma screening potential.
In contrast to Ref.~\cite{Kravchuk.2014.PRC},
we calculate pure Coulomb potential $v_{C} (r)$ on the basis of Eqs.~(\ref{eq.analysis.3.2})--(\ref{eq.analysis.3.4})
(here, Coulomb potential in the nuclear region at $r < R_{C}$ is different from corresponding potential in Eq.~(7) in Ref.~\cite{Kravchuk.2014.PRC}),
and we use nuclear potential $v_{N}(r)$ in Eq.~(\ref{eq.analysis.3.2})
(we set $l=0$).
In definition of the screening part of potential we follow Ref.~\cite{Kravchuk.2014.PRC} and use
(see Eqs.~(10), (11) in that paper):
\begin{equation}
\begin{array}{llllll}
  H(r) = E_{12}\, h(x), &
  x = \displaystyle\frac{r}{a_{12}},
\end{array}
\label{eq.screening.1.2}
\end{equation}
where
\begin{equation}
\begin{array}{llllll}
  h(x) = b_{0} + b_{2}\, x^{2} + b_{4}\, x^{4} + \ldots.
\end{array}
\label{eq.screening.1.3}
\end{equation}
At $Z_{1} / Z_{2} = 1$ parameters $b_{0}$, $b_{2}$, $b_{4}$ are derived in Ref.~\cite{Kravchuk.2014.PRC} as
\begin{equation}
\begin{array}{llllll}
  b_{0} = 1.0573, & b_{2} = -0.25, & b_{4} = 0.0394.
\end{array}
\label{eq.screening.1.4}
\end{equation}
Other parameters are
(see Eqs.~(2), (4) in Ref.~\cite{Kravchuk.2014.PRC})
\begin{equation}
\begin{array}{llllll}
  a_{e} = \Bigl( \displaystyle\frac{3}{4\pi\, n_{e}} \Bigr)^{1/3}, &
  a_{j} = Z_{j}^{1/3}\, a_{e},
\end{array}
\label{eq.screening.1.5}
\end{equation}
\begin{equation}
\begin{array}{llllll}
  a_{12} = \displaystyle\frac{a_{1} + a_{2}}{2}, &
  E_{12} = \displaystyle\frac{Z_{1} Z_{2}\, e^{2}}{a_{12}}, &
\end{array}
\label{eq.screening.1.6}
\end{equation}
where
$n_{e}$ is concentration of electrons.

On the basis of Eqs.~(\ref{eq.analysis.2.6}) we calculate concentration of nuclei $n_{A}$ at the studied density as
%
%
\begin{equation}
\begin{array}{llllll}
  \isotope[12]{C} + \isotope[12]{C}, &
  \rho_{0} = 6 \cdot 10^{9}\, \displaystyle\frac{\mbox{g}} {\mbox{\rm cm}^{3}}: &
  n_{A} =
  3.\: 014\: 18 \cdot 10^{-7}\; \mbox{\rm fm}^{-3}.
\end{array}
\label{eq.screening.1.7}
\end{equation}
This can be understood as each nucleus \isotope[12]{C} is in volume with size about 200~fm.
From here we find concentration of electrons
%
\begin{equation}
\begin{array}{llllll}
  n_{e} = Z \cdot n_{A}, &

  n_{e} =
  1.\: 808\: 51 \cdot 10^{-6}\; \mbox{\rm fm}^{-3}
%
\end{array}
\label{eq.screening.1.8}
\end{equation}
and from Eqs.~(\ref{eq.screening.1.3})--(\ref{eq.screening.1.4}) we obtain
%
\begin{equation}
\begin{array}{llllll}
  a_{12} = 92.\: 522\: 41\; \mbox{\rm fm}.
\end{array}
\label{eq.screening.1.9}
\end{equation}
As it is indicated in Ref.~\cite{Kravchuk.2014.PRC}, Eq.~(\ref{eq.screening.1.3}) should be used at $x \ll 2$.
We estimate that this condition is fulfilled in the full region of study of reaction
$\isotope[12]{C} + \isotope[12]{C}$ at the chosen density,
so we use Eq.~(\ref{eq.screening.1.3}) in description of the screening part of the potential.

\begin{figure}[htbp]
\centerline{\includegraphics[width=88mm]{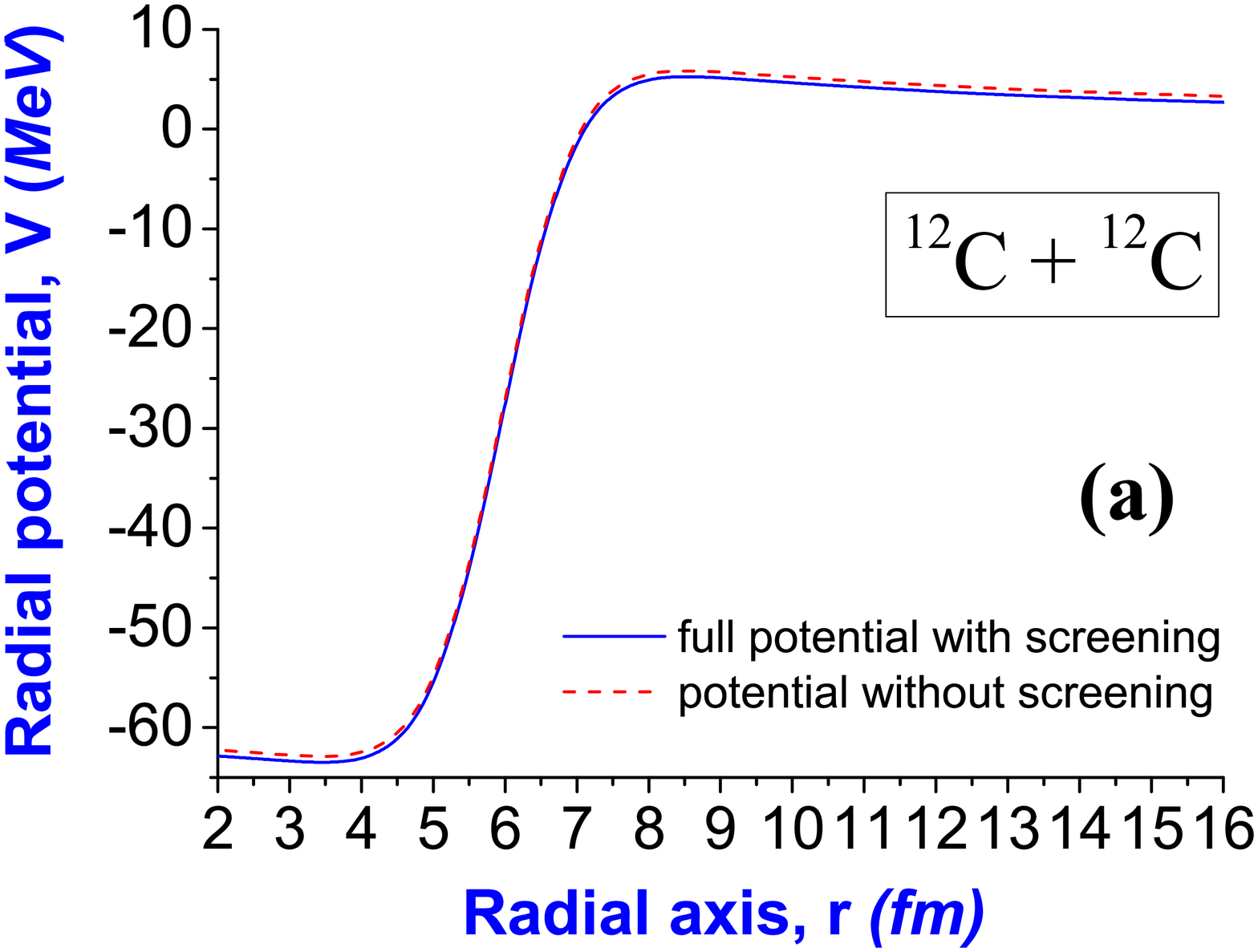}
\hspace{-3mm}\includegraphics[width=88mm]{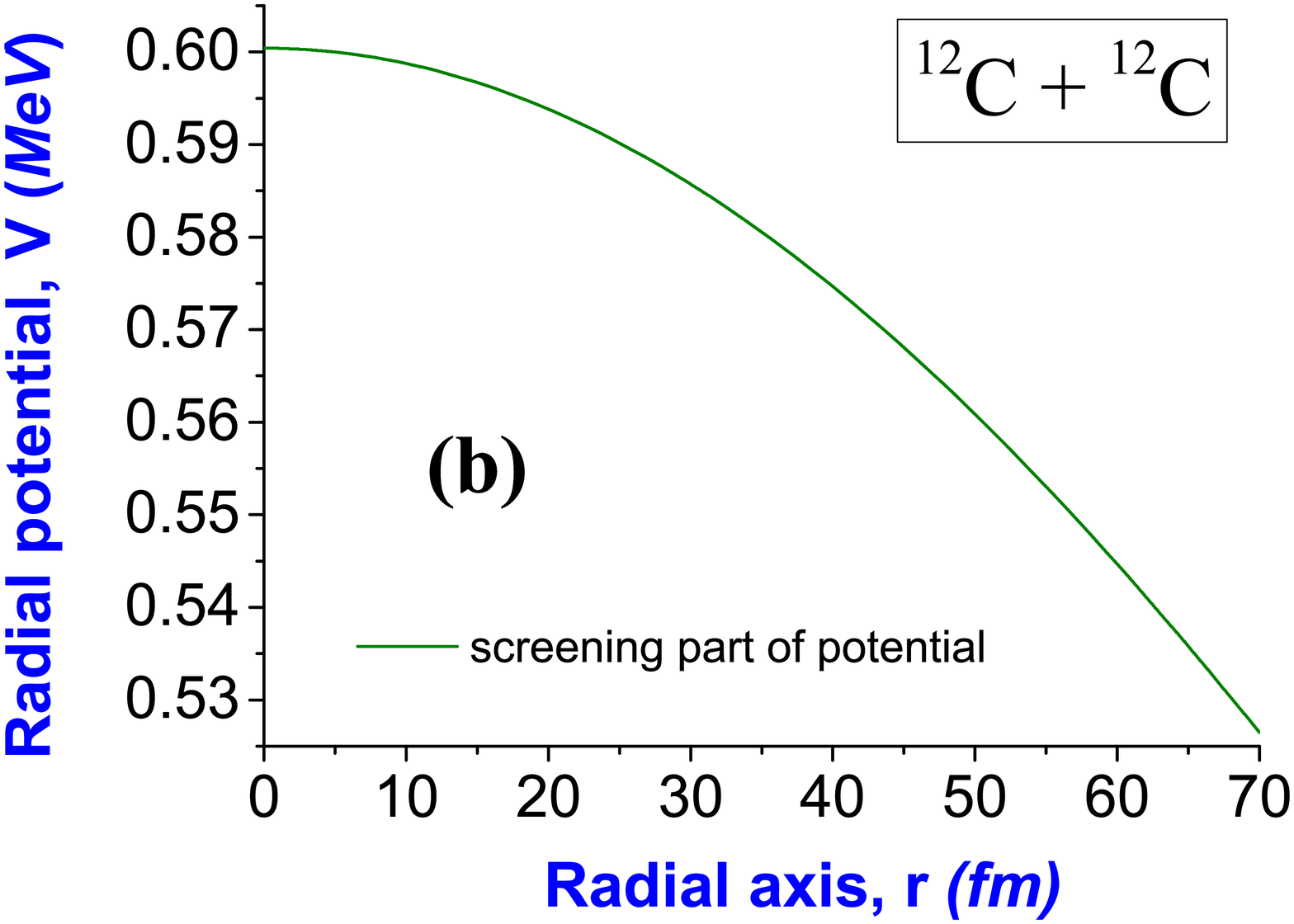}}
\caption{\small (Color online)
Potential of interaction between two nuclei \isotope[12]{C}
with inclusion of screening in comparison with the same potential without screening (a)
and part of potential describing screening (b)
[potential and parameters are defined in Eq.~(\ref{eq.analysis.3.1})--(\ref{eq.analysis.3.3})
screening part of the potential is defined in Eqs.~(\ref{eq.screening.1.1})--(\ref{eq.screening.1.3})].
\label{fig.potential_screening}}
\end{figure}

Potential of interactions with taking into account screening, calculated by such an approach, is shown in Fig.~\ref{fig.potential_screening}.
%
%
From this figure one can see that the screening does not change the potential much at the density of matter under consideration.
So, one can suppose that the screening does not influence essentially on results of quasibound states and energies above.
But, energies of zero point vibrations are essentially smaller than quasibound energies,
and one can suppose that energy spectrum of zero-point vibrations will be changed after inclusion of plasma screening to calculations.

In Tabl.~\ref{table.screening.1} values of energies of zero-point vibrations calculated for reaction $\isotope[12]{C} + \isotope[12]{C}$
are presented taking into account screening and without it.
%
\begin{table}
\begin{center}
\begin{tabular}{|c|c|c|c|c|c|c|l|l|l|l|} \hline
  No.    & \isotope[12]{C} + \isotope[12]{C} with screening
         & \isotope[12]{C} + \isotope[12]{C} without screening
         \\ \hline
 1  &   0.209619238476954 & 
        0.517434869739479 
        \\

 2  &  0.267334669338677 & 
       0.536673346693387 
       \\

 3  & 0.344288577154309 & 
      0.546292585170341 
      \\

 4  & 0.478957915831663 & 
      0.555911823647295 
      \\

 5  & 0.786773547094188 & 
      0.575150300601202 
      \\

 6 & 2.21042084168337 & 
     0.594388777555110 
     \\

 7 & 3.83607214428858 & 
     0.613627254509018 
     \\

 8 & &
     0.642484969939880 
     \\

 9  & &
      0.680961923847695 
      \\

 10  & &
      0.729058116232465 
      \\

 11  & &
       0.806012024048096 
       \\

 12 & &
     0.911823647294589 
     \\

 13 & &
     1.11382765531062 
     \\

 14 & & 
      2.76833667334669 
      \\

15 & --- &
   4.08617234468938 
   \\
\hline
\end{tabular}
\end{center}
\caption{%
Energies for zero-point vibrations $E_{\rm zero}^{\rm (mir)}$ (values are presented in MeV, below 5 MeV)
calculated for reaction $\isotope[12]{C} + \isotope[12]{C}$.
}
\label{table.screening.1}
\end{table}
From this table one can see that energy spectrum of zero-point vibrations for reaction $\isotope[12]{C} + \isotope[12]{C}$
is changed essentially after taking into account plasma screening.

\section{Conclusions and perspectives
\label{sec.conclusion}}

%
Question of conditions needed for the most probable formation of compound nuclei (as the first stage needed for synthesis of more heavy elements)
in pycnonuclear reactions in compact stars is investigated in this paper.
The method is developed on the basis of formalism of the multiple internal reflections, previously constructed for study of quantum processes and phenomena with high precision and testing in nuclear decays~\cite{Maydanyuk.2011.JPS,Maydanyuk.2000.UPJ,Maydanyuk.2002.JPS,Maydanyuk.2006.FPL},
nuclear captures by nuclei~\cite{Maydanyuk.2015.NPA,Maydanyuk_Zhang_Zou.2017.PRC},
as well as in problems of quantum cosmology where the idea of tunneling was investigated \cite{Maydanyuk.2011.EPJP}.
In this paper we continue investigations of pycnonuclear reactions with isotopes of Carbon,
started in Ref.~\cite{Maydanyuk_Shaulskyi.2022.EPJA} for $\isotope[12]{C} + \isotope[12]{C}$.
Conclusions of our analysis are the following.

\begin{itemize}
\item
In this research, pycnonuclear processes are studied taking the nuclear part of potential of interactions between nuclei into account.
Requirement of continuity of quantum flux (describing pycnonuclear reaction on the basis of quantum mechanics) gives
appearance of new states, in which compound nuclear system of \isotope[2X]{Mg} is formed with the highest probability
(see Fig.~\ref{fig.3.4}).
Following to logic in Refs.~\cite{Maydanyuk_Shaulskyi.2022.EPJA,Maydanyuk.2015.NPA,Maydanyuk_Zhang_Zou.2017.PRC},
we call such states as \emph{quasi-bound states in pycnonuclear reactions}.
Note that these states have not been studied yet by other researchers in study of synthesis of elements in stars.

\item
As shown in Fig.~\ref{fig.3.4},
probability of formation of compound nuclear system in quasibound states is essentially higher than
probability of formation of this system in states of zero-point vibrations
studied by Zel'dovich~\cite{Zeldovich.1965.AstrJ} and followers of that idea.
Therefore, synthesis of more heavy nuclei of Magnesium from isotopes of Carbon is essentially more probable in quasibound states
than in states of zero-point vibrations.
This leads to revision (reconsideration) of pictures of formation of heavy elements in compact stars,
to use quasibound states as the basis for synthesis.
This leads to the essential changes in estimation of the rates of pycnonuclear reactions in stars. 
One can note the perspective to test the method presented in this paper
on the basis of experimental measurements in Ref.~\cite{Fang.2017.PRC}.

\item
Only the first quasibound energies for
$\isotope[10]{C} + \isotope[10]{C}$,
$\isotope[12]{C} + \isotope[12]{C}$,
$\isotope[24]{C} + \isotope[24]{C}$
(see Tabl.~\ref{table.2}) are smaller than the barrier maximums for these nuclear systems
(see Tabl.~\ref{table.analysis.3.1}).
So, at such energies
the compound nuclear systems have barriers which prevent their decays going through tunneling phenomenon.
At such energies the compound nuclear systems are the most probable and the most longer lived.
These systems are transformed to new synthesized isotopes of Magnesium \isotope[20]{Mg}, \isotope[24]{Mg} and \isotope[48]{Mg} with large probabilities.
There is a simple way to estimate half-lives of these obtained more heavy nuclei using Gamow's approach or the method of Multiple internal reflections for higher precision.
Note that other approaches cannot estimate the quasibound energies needed for prediction of synthesis of more stable nuclear systems by such a way described above.
At the same time, the method of Multiple internal reflections calculates such energies with high precision, providing also tests to check calculations.
But, analysis of binding energies for the obtained isotopes of Magnesium shows that only \isotope[24]{Mg} will be stable after synthesis.



\item
At the first time influence of plasma screening on quasibound states and states of zero-point vibrations in pycnonuclear reactions has been studied.
It is found that energy spectrum of zero-point vibrations is changed essentially after taking into account plasma screening
(see Tabl.~\ref{table.screening.1} for reaction $\isotope[12]{C} + \isotope[12]{C}$).
\end{itemize}

\section*{Acknowledgements
\label{sec.acknowledgements}}

S.~P.~M. thanks the Wigner Research Centre for Physics in Budapest for warm hospitality and support.
Authors are highly appreciated to
Prof.~V.~S.~Vasilevsky for fruitful discussions concerning to mechanisms of formation of the compound nuclear system and scattering of nuclei,
peculiarities of the method of multiple internal reflections,
Prof.~A.~G.~Magner for fruitful discussions concerning to the aspects of nuclear matter in conditions of compact stars and in Earth,
Prof.~V.~Ambrus for fruitful discussions concerning to physical issues of astrophysical $S$-factors and applicability of quantum methods in stars.
%
%
G.~W. and S.~P.~M. thank the support of OTKA grant K138277.

\end{document}